\documentclass[11pt,a4paper]{article} 

\usepackage{ajtex}
\usepackage{ajtex-defs}

\usepackage{setspace}
\usepackage{footnote}
\makesavenoteenv{tabular}
\makesavenoteenv{table}

\title{Second order Galilean fluids \& Stokes' law}

\author[a]{Nabamita Banerjee}
\email{nabamita@iiserpune.ac.in}

\author[a]{Sayali Atul Bhatkar}
\email{sayali.bhatkar@students.iiserpune.ac.in}

\author[b]{Akash Jain}
\email{akash.jain@durham.ac.uk}

\affiliation[a]{Dept.\ of Physics, Indian Institute of Science Education and
  Research (IISER), Pune 411008, Maharashtra, India}

\affiliation[b]{Centre for Particle Theory \& Dept.\ of Mathematical Sciences,
  Durham University, Durham DH1 3LE, United Kingdom}

\abstract{ We study the second derivative effects on the constitutive relations
  of an uncharged parity-even Galilean fluid using the null fluid
  framework. Null fluids are an equivalent representation of Galilean fluids in
  terms of a higher dimensional relativistic fluid, which makes the Galilean
  symmetries manifest and tractable. The analysis is based on the offshell
  formalism of hydrodynamics. We use this formalism to work out a generic
  algorithm to obtain the constitutive relations of a Galilean fluid up to
  arbitrarily high derivative orders, and later specialise to second
  order. Finally, we study the Stokes' law which determines the drag force on an
  object moving through a fluid, in presence of certain second order terms. We
  identify the second order transport coefficients which leave the drag force
  invariant.}

\preprint{DCPT-17/37}

\newcommand\bref[1]{(\ref{#1})}

\begin{document}
\maketitle

\section{Introduction}

Hydrodynamics is the universal low energy effective description of a finite
temperature field theory around its thermodynamic equilibrium. A fluid
configuration is described by certain fluid variables, typically chosen to be a
normalised velocity, temperature and chemical potentials associated with any
internal symmetries. Dynamical equations for these variables are provided by the
conservation of energy-momentum and any additional charges associated with the
internal symmetries. By the virtue of being a low energy description, the length
scales over which the fluid variables vary are taken to be large compared to the
mean free path of the system. Thus, fluid energy-momentum tensor and charge
currents admit a perturbative expansion in terms of derivatives of fluid
variables, known as the fluid constitutive relations. At any given order in the
derivative expansion, the constitutive relations contain all the possible tensor
structures made out of fluid variables and their derivatives, multiplied with
arbitrary functions of temperature and chemical potentials. These coefficients
are known as the transport coefficients of the fluid. It is well known that if
the dissipative hydrodynamics is truncated only at first derivative order with
shear and bulk viscosity coefficients, there are always linearised fluctuations
for which the wave-front speed is superluminal~\cite{Hiscock:1987zz,
  Hiscock:1986zz, Hiscock:1983zz} and thus the causal structure of the theory is
broken. Thus within the hydrodynamic framework, it was noted long ago by Muller,
Israel and Stewart~\cite{Israel:1979wp, Israel:1976tn, Muller:1967zza} that one
needs to go beyond the first derivative order and add specific second order
terms to address the issue of causality. Since a causal system of second-order
hydrodynamic equations is required in many situations, such as numerical
simulations~\cite{Song:2008hj, Luzum:2008cw}, a more systematic and detailed
analysis of second order fluid transport has since been
performed~\cite{Bhattacharyya:2012nq}.

Although relativistic hydrodynamics is extremely useful at various physical
fronts, non-relativistic hydrodynamics also enjoys an active interest in the
physics community. On our day to day energy scales, the world is governed by
non-relativistic physics. In this sense, a non-relativistic fluid can be thought
of as an effective version of a relativistic fluid under a non-relativistic
(large speed of light) limit. Hence, it is natural to expect that the associated
constitutive relations, obtained as an effective description of a relativistic
theory, may contain new terms which were not considered in the coarse grained
description of relativistic hydrodynamics. This is apparent even at the first
derivative order wherein the non-relativistic fluid contains an additional
transport coefficient called the thermal conductivity.

Conventionally, non-relativistic hydrodynamics is studied by writing down a
fluid theory whose fundamental symmetry group is Galilean as opposed to
Poincar\'e for relativistic hydrodynamics. However, as the Galilean symmetry
treats time and space coordinates on a different footing, the analysis becomes
much more cumbersome. In a series of papers~\cite{Banerjee:2015uta,
  Banerjee:2015hra, Jain:2015jla, Banerjee:2016qxf}, we devised a new mechanism
to study non-relativistic hydrodynamics. We constructed a tweaked relativistic
fluid in one higher dimension, called a null fluid, and showed that it is
equivalent to a Galilean fluid in all respects. We also provided a simple
dictionary, using the well established procedure of null
reduction~\cite{Rangamani:2008gi}, to translate between the null fluid and
conventional Galilean fluid languages. Apart from representing Galilean fluids
in a familiar and intuitive relativistic framework, null fluids also allow us to
directly import the well developed results and machinery of relativistic
hydrodynamics into Galilean fluids. For instance, we have successfully employed
the null fluid framework to study first order (anomalous) charged Galilean
hydrodynamics in~\cite{Banerjee:2015uta, Banerjee:2015hra}, using both the
second law of thermodynamics and equilibrium partition
functions~\cite{Banerjee:2012iz,Jensen:2012jh} to determine physical constraints
on the transport coefficients. We have also introduced the offshell formalism
for relativistic hydrodynamics~\cite{Loganayagam:2011mu} into Galilean
hydrodynamics, and used it to study the influence of anomalies on the Galilean
fluid constitutive relations~\cite{Jain:2015jla} following the corresponding
development in relativistic hydrodynamics (see~\cite{Son:2009tf,
  Loganayagam:2011mu, Jensen:2013kka} and references therein). Recently,
following the analysis of relativistic superfluids in
\cite{Bhattacharya:2011tra, Bhattacharyya:2012xi, Jain:2016rlz}, we studied
first order (anomalous) Galilean superfluids in~\cite{Banerjee:2016qxf}. These
results were used to study surface transport in Galilean superfluids
in~\cite{Armas:2016xxg}.

In this paper, our aim is to perform a complete analysis of second order
Galilean hydrodynamics using null fluids. As we stated above, causality demands
the inclusion of second order transport coefficients in relativistic
hydrodynamics, so it is natural to expect an impression of these to propagate
into non-relativistic hydrodynamics as well. Moreover, since null fluids are a
relativistic system, we are also required to include the second order terms on
grounds of consistency with causality. We focus on the parity-even uncharged
case for computational simplicity, although the generalisation of our results,
albeit computationally involved, should be straight forward. We start by
outlining a generic algorithm to work out the constitutive relations of a
Galilean fluid up to arbitrary orders in derivative expansion in offshell
formalism. This is based on a recent classification scheme for the entire
relativistic hydrodynamic transport presented in~\cite{Haehl:2014zda,
  Haehl:2015pja, Jain:2016rlz}. Later, we use this algorithm to study the second
order Galilean fluids. We find that in addition to the pressure at ideal order,
and bulk viscosity, shear viscosity and thermal conductivity at first order,
there are 25 coefficients at second order. 5 of them are hydrostatic, i.e.\ they
govern the equilibrium configuration of the fluid. 9 are dissipative, i.e.\ they
are responsible for the production of entropy during dynamical processes, while
the remaining 11 quantify dynamical processes which do not cause any dissipation.

To explore the physical significance of these second order transport
coefficients, we study the effect of some representative terms on the Stokes'
law. It is a famous hydrodynamic equation with numerous applications in physics
and even in biology, that determines the drag force experienced by a body when
moving through a fluid. If we acknowledge that there is an underlying causal
relativistic theory behind our non-relativistic fluid, of which our
non-relativistic fluid is just a low-energy description, this simple law will
get modified by inclusion of the appropriate second order effects. In this
article, we shall explore how some of the second order transport coefficients
appearing in non-relativistic hydrodynamics might affect the Stokes' law. We
also identify a class of second order terms, which leave the Stokes' law invariant.

We should comment that in this paper we use the terms ``non-relativistic fluid''
and ``Galilean fluid'' interchangeably. In principle, a Galilean fluid is
defined as the most generic fluid which obeys Galilean symmetries. In this
sense, a non-relativistic fluid is a special kind of Galilean fluid which
follows by taking a non-relativistic limit of a relativistic system. This is to
say that there might be some additional constraints on the Galilean fluid,
following from the requirement that it should follow under a non-relativistic
limit. In our previous work however (see section~5 of~\cite{Banerjee:2016qxf}),
we argued that there are no such additional constraints and that every Galilean
fluid can in fact be obtained via a non-relativistic limit.

The paper is organised as follows. We start \cref{sec:nullfluids} with a brief
review of null fluids and offshell formalism. We outline the generic algorithm
to construct the null fluid constitutive relations at arbitrarily high
derivative orders in \cref{sec:classification}, and illustrate how first order
constitutive relations fit into this scheme in \cref{firsorderresults}. In
\Cref{sec:2ndNull}, we use this algorithm to work out the second order
constitutive relations of a Galilean fluid. For readers not interested in the
computational details, the results have been summarised in
\cref{tab:2nd-order-hydrostatic,tab:2nd-order-hydrostatic-N,tab:1der-C0,tab:0der-C1,tab:0der-C1-N}. In
\cref{sec-redn}, we review the translation of null fluid constitutive relations
into the conventional Galilean notation, and use it to obtain the constitutive
relations of a Galilean fluid up to second derivative order. The final results
gave been given in \cref{tab:NC-r,tab:NC-e,tab:NC-t}.  We explore how these
second order terms might affect the well known Stokes' law in
\cref{sec:application}, exploring the physical significance of our
results. Finally, we close the paper with some discussion in
\cref{sec:discussion}.

\section{Crash course in null fluids}\label{sec:nullfluids}

In~\cite{Banerjee:2015hra} we proposed ``null fluids'' as a new viewpoint of
Galilean fluids. The main benefit of working in this formalism is that it is
effectively a ``relativistic embedding'' of a Galilean fluid into one higher
dimension.  This enables us to import the preexisting relativistic machinery and
intuition into Galilean hydrodynamics. In this section we collect some of the
results about null fluids which we use throughout this paper. The discussion is
self contained, albeit brief. For more details, the reader is encouraged to
refer to our previous papers, especially~\cite{Banerjee:2015hra,
  Banerjee:2016qxf, Jain:2016rlz}.

\subsection{Null backgrounds and null fluids}

Simply put, hydrodynamics is the study of long wavelength fluctuations of a
quantum system on top of some slowly varying background fields. In this sense,
the rules of hydrodynamics are governed by the background we start with. For
example, consider a $(d+2)$-dimensional background with a metric $g_{\sM\sN}$
and an invariance under diffeomorphisms which act on the metric as
\begin{equation}
  \d_{"scX} g_{\sM\sN} = \lie_{"c}g_{"sM"sN} = \N_\sM \c_\sN + \N_\sN \c_\sM.
\end{equation}
Here $"scX = \{\c^\sM\}$ are some arbitrary parameters and $\N_{"sM}$ is the
covariant derivative associated with the Levi-Civita connection
\begin{equation}
  \G^{"sR}{}_{"sM"sN}
  = \half g^{"sR"sS}\lb \dow_{"sM}g_{"sN"sS}
  + \dow_{"sN} g_{"sM"sS} - \dow_{"sR} g_{"sM"sN}
  \rb.
\end{equation}
Hydrodynamics on this background corresponds to a usual (uncharged) relativistic
fluid. To study null fluids however, we need to tweak this background by
introduction of a vector field $"scV = \{V^\sM\}$ which is null
$V^\sM V_\sM = 0$, covariantly constant $\N_\sM V^\sN = 0$ and is an isometry
\begin{align}
  \d_{"scV} g_{\sM\sN}
  = \lie_{V}g_{\sM\sN} = \N_\sM V_\sN + \N_\sN V_\sM = 0.
\end{align}
We call these \emph{null backgrounds}. They provide a natural ``embedding'' for
Galilean (Newton-Cartan) backgrounds into a relativistic spacetime of one higher
dimension. We will only be interested in fluctuations that respect the
symmetries of the background, i.e.\ they must transform appropriately under
diffeomorphisms and admit $\d_{"scV} = 0$. Few key points to note about null
backgrounds: the Riemann curvature tensor
\begin{equation}\label{RiemannDef}
  R^{"sM}{}_{"sN"sR"sS} = 2\dow_{["sR}\G^{"sM}{}_{"sS]"sN} +
  2 "G^{"sM}{}_{"sT["sR} "G^{"sT}{}_{"sS]"sN},
\end{equation}
has vanishing contractions with $"scV$:
$R^{"sM}{}_{"sN"sR"sS} V^{"sN} = 2\N_{["sR}\N_{"sS]}V^{"sM} = 0$. Furthermore,
we have a consistency condition on $"scV$:
$H_{"sM"sN} \equiv 2 \dow_{["sM}V_{"sN]} = 2 \N_{["sM}V_{"sN]} = 0$. From a
mathematical standpoint, perhaps it is more natural to introduce a torsion
tensor $"rmT^{"sR}{}_{"sM"sN}$, in presence of which this condition becomes
$V_{"sR}"rmT^{"sR}{}_{"sM"sN} = H_{"sM"sN}$ and lifts the constraint from
$"scV$. However, in interest of simplicity, we will work with torsion-less null
backgrounds and comment on the ``unnaturalness'' as it arises.

Given the diffeomorphism symmetry, Noether theorem implies that our theory has
an energy-momentum tensor $T^{\sM\sN}$ in its spectrum. The respective
conservation law is given as
\begin{equation}\label{eom_gsf}
  \N_\sM T^{\sM\sN} = 0.
\end{equation}
On null backgrounds we must further require that $"d_{"scV}T^{"sM"sN} = 0$.
Consequently, $T^{\sM\sN}$ is only defined up to terms proportional to
$V^{"sM}V^{"sN}$ as $\N_{"sM}(\# V^{"sM}V^{"sN}) = V^{"sN}\d_{"scV}\# = 0$. We
will extensively use this freedom to ignore terms proportional to
$V^{"sM}V^{"sN}$ in $T^{"sM"sN}$ throughout this paper.

Having $(d+2)$ independent components, the conservation law \bref{eom_gsf} can
provide dynamics for a ``fluid theory'' formulated in terms of arbitrary $(d+2)$
variables. We choose these to be a normalised null \emph{fluid velocity} $u^\sM$
(with $u^\sM V_\sM = -1$, $u^\sM u_\sM = 0$), a \emph{temperature} $T$ and a
\emph{mass chemical potential} $\mu_{m}$, collectively known as the
\emph{hydrodynamic fields}. We sometimes work with a scaled mass chemical
potential $"vp = "m_m/T$. On a null background, we must further demand these
fields to be compatible with $"scV$, i.e.
$"d_{"scV}u^{"sM} = "d_{"scV} T = "d_{"scV} "m_m = 0$.  In general, $T^{"sM"sN}$
can arbitrarily depend on the hydrodynamic and background fields. But
fortunately in hydrodynamics, we are only interested in low energy fluctuations
of the hydrodynamic fields and a slowly varying background. This allows us to
treat derivatives as a perturbation parameter. A null fluid is therefore
completely characterised by a covariant expression for $T^{\sM\sN}$ in terms of
$g_{\sM\sN}$, $V^{"sM}$, $u^\sM$, $T$, $\mu_m$ and their derivatives, truncated
to a desired derivative order, known as the \emph{null fluid constitutive
  relations}. At any given order, the constitutive relations can admit some
tensor structures made out of the constituent fields and their derivatives
called \emph{data}, multiplied with arbitrary functions of $T$ and $"m_m$ called
transport coefficients. To this end, hydrodynamics is just a combinatorial
exercise of enlisting all the possible tensor structures at a given order in
derivatives.

To make things more interesting, we need to realise that fluids are
thermodynamic systems. We have already imposed the first law of thermodynamics
(conservation of energy) implicitly in form of the conservation laws
\bref{eom_gsf}. In addition, they are also required to satisfy a version of the
\emph{second law of thermodynamics}. It states that there must exist an
\emph{entropy current} $J_S^\sM$ whose divergence is positive semi-definite
everywhere, i.e.,
\begin{equation}\label{onshell2ndlaw} 
  \N_\sM J_S^\sM = "D \geq 0,
\end{equation}
as long as the fluid is thermodynamically isolated (i.e.\ conservation laws
\cref{eom_gsf} are satisfied). Note that like $T^{"sM"sN}$, the entropy current
$J^{"sM}_S$ is only defined up to terms proportional to $V^{"sM}$. In null
hydrodynamics, our goal is to find the most generic constitutive relations
$T^{\sM\sN}$ and some associated $J_S^\sM$ and $"D$ order by order in derivative
expansion, such that \cref{onshell2ndlaw} is satisfied for all thermodynamically
isolated fluids. As innocuous as this statement sounds, the complexity involved
drastically increases as we increase the derivative order.  There is however, an
equivalent but much neater way to work out these constitutive relations, called
the \emph{offshell formalism}~\cite{Loganayagam:2011mu}. In the next subsection,
we adapt this formalism to null fluids. 

\subsection{Offshell formalism}

As it turns out, most of the trouble while implementing \cref{onshell2ndlaw}
roots from the fact that it needs to only be imposed on the solutions of the
equations of motion, i.e.\ onshell.  We can relax this condition by coupling the
fluid to an external momentum source $P^\sM_{ext}$, so that the conservation law
\bref{eom_gsf} is no longer satisfied. Having done that, \cref{onshell2ndlaw}
gets modified with an arbitrary combination of $P^\sM_{ext}$ giving us
\begin{equation}\label{offshell_2ndlaw}
  \N_\sM J_S^\sM + \b_\sM P^\sM_{ext}
  = \N_\sM J_S^\sM + \b_\sN \N_\sM T^{\sM\sN}
  = "D \geq 0,
\end{equation}
where $"scB = \{\b^\sM\}$ is an arbitrary vector multiplier. This equation is
referred to as the \emph{offshell second law of thermodynamics}. It can be
rewritten into a more useful form by defining a \emph{free energy current}
$G^{"sM}$
\begin{equation}
  -\frac{G^\sM}{T} = N^\sM = J_S^\sM + T^{\sM\sN} \b_\sN.
\end{equation}
\Cref{offshell_2ndlaw} now turns into a conservation equation for free energy
\begin{equation}\label{adiabaticity}
  \N_\sM N^\sM = \half T^{\sM\sN} \d_{"scB} g_{\sM\sN} + "D,
  \qquad "D \geq 0.
\end{equation}
Here $\d_{"scB} g_{\sM\sN}$ is of course $2\N_{("sM}"b_{"sN)}$. Recall that the
hydrodynamic fields $u^\sM$, $T$ and $\mu_m$ were some arbitrary fields chosen
to describe the fluid. Like in any field theory, they are permitted to admit an
arbitrary redefinition among themselves without changing the physics. This huge
amount of freedom can be fixed by explicitly choosing
\begin{equation}\label{offshell-hydro-frame}
  u^\sM = -\frac{\b^\sM}{V_\sM \b^\sM} + \frac{\b^\sR \b_\sR V^\sM}
  {2{(V_\sN \b^\sN)}^2}, \qquad
  T = -\frac{1}{V_\sM \b^\sM}, \qquad
  \mu_m = \frac{\b^\sM "b_\sM}{2{(V_\sN \b^\sN)}^2},
\end{equation}
or conversely
\begin{equation}
  \b^\sM = \frac{1}{T} \lb u^\sM - \mu_m V^\sM \rb.
\end{equation}
It follows that $"b^{"sM}$ can be understood as a rewriting of the conventional
hydrodynamic fields $u^\sM$, $T$ and $\mu_m$. Although this is a very convenient
``frame'' choice for our analysis, it might not be the most useful one for
applications. But once we have obtained a consistent set of constitutive
relations, we can always field transform to any desired frame.

To agree with the second law of thermodynamics, we need to find the most generic
$T^{"sM"sN}$ written in terms of $"b^{"sM}$, $g_{"sM"sN}$ and $V^{"sM}$ which
satisfies \cref{adiabaticity} for some $N^{"sM}$ and $"D$. There is one minor
subtlety to keep in mind though: $T^{"sM"sN}$ found this way will also contain
information about the external sources $P^{"sN}_{ext} = \N_\sM
T^{\sM\sN}$. Therefore, in the end we must identify the constitutive relations
which are related to each other up to combinations of equations of motion or
their derivatives. Generically, equations of motion determine the ``time''
derivative of the fundamental fields, so without any loss of generality we can
use them to eliminate
$u^{"sM}\N_{"sM}"b_{"sN} + u^{"sM}\N_{"sN}"b_{"sM} = u^{"sM}"d_{"scB}g_{"sM"sN}$
from our constitutive relations. Consequently, we will only be interested in the
constitutive relations $T^{"sM"sN}$ which are independent of
$u^{"sM}"d_{"scB}g_{"sM"sN}$.

\subsection{Classification of constitutive relations}
\label{sec:classification}

We need to find the most generic solutions to \cref{adiabaticity} up to a given
order in derivatives. We could take the ``brute-force'' approach wherein we plug
in the most generic expressions for $T^{"sM"sN}$ and $N^{"sM}$ up to a given
derivative order, and find constraints arising from $"D\geq 0$. These
constraints can be fairly complicated at higher derivative orders and crucially
depend on the choice of basis for the tensor-decomposition of $T^{"sM"sN}$ and
$N^{"sM}$. But we can do better: we can write down a ``solution generating
algorithm'' following~\cite{Haehl:2014zda, Jain:2016rlz} which will work at
arbitrarily high derivative orders, as we now outline. We will then go on to
apply this algorithm to second order null fluids in \cref{sec:2ndNull}. The
discussion in this subsection parallels \cite{Jain:2016rlz}.

First and foremost we consider trivial solutions of \cref{adiabaticity} which do
not contribute to the constitutive relations:
\begin{itemize}[leftmargin=1em]
\item \textbf{Entropy transport (Class S)}: These are solutions of
  \cref{adiabaticity} of the kind $N^\sM = N^\sM_{"rmS}$,
  $T^{\sM\sN}_{"rmS} = 0$ such that $\N_\sM N^\sM_{"rmS} = "D_{"rmS}$ is a
  quadratic form. They contain, for example,
  $N^\sM_{"rmS} = \N_{\sN} X^{[\sM\sN]}$ for an arbitrary antisymmetric tensor
  $X^{[\sM\sN]}$ or $N^\sM_{"rmS} = "cS V^\sM$ for an arbitrary scalar $"cS$,
  for both of which $"D_{"rmS} = 0$. Examples for $"D_{"rmS} \neq 0$ are
  slightly complicated but can be obtained after some effort. These solutions
  correspond to the transport of entropy in a fluid
  ${(J^\sM_S)}_{"rmS} = N^\sM_{"rmS}$, without any transport of energy-momentum.
\end{itemize}
Note that if $T^{\sM\sN}$ is a solution to \cref{adiabaticity} for some free
energy current $N^{"sM}$, then instead the free energy current
$N^\sM + N^\sM_{"rmS}$ would also do. Therefore to satisfy the second law of
thermodynamics for a given set of constitutive relations $T^{\sM\sN}$, we can
choose any entropy current from the equivalence class
$J^\sM_S \sim J^\sM_S + N^\sM_{"rmS}$. In a strict sense therefore, Class S
solutions are not really ``hydrodynamic''. They merely parametrise the multitude
of entropy currents which might satisfy the second law for a given set of
constitutive relations.

Getting these redundancies out of the way, we can now focus on the actual
physical solutions. We broadly split the constitutive relations into two
sectors: \emph{hydrostatic} and \emph{non-hydrostatic} based on the tensor
structures that go into making them. Non-hydrostatic tensor structures are those
which contain at least one instance of $\d_\scB g_{"sM"sN}$ or its
derivatives. On the other hand, hydrostatic tensor structures are the largest
collection of independent tensor structures with no non-hydrostatic linear
combination. The terminology is based on the concept of \emph{equilibrium}: a
background is said to admit a hydrostatic configuration if it has a timelike
isometry $"scK = \{K^{"sM}\}$. On such a background, a hydrostatic configuration
is given by $"b^{"sM} = K^{"sM}$ which trivially satisfies the equations of
motion. By definition therefore, hydrostatic constitutive relations are the only
ones that survive in a hydrostatic configuration and govern the equilibrium
physics.

Spoiler alert! The second law imposes strict constraints in the hydrostatic
sector allowing for only specific combinations of tensor structures to appear,
while in the non-hydrostatic sector it merely gives a few inequalities at the
first order in derivatives and none thereafter
\cite{Bhattacharyya:2013lha,Bhattacharyya:2014bha}.

\begin{itemize}[leftmargin=1em]
\item \textbf{Hydrostatic transport (Classes H$_{"bS}$, H$_{"bV}$ and A):} These
  are solutions of \cref{adiabaticity} made purely out of hydrostatic tensor
  structures. They have $"D = 0$ and are completely characterised by a free
  energy current of the form
  \begin{equation}
    N^\sM = \lb "cN "b^\sM + "Q_{"cN}^\sM \rb + "bbN^\sM.
  \end{equation}
  Here $"cN$ is the most generic hydrostatic scalar, while $"Q_{"cN}^\sM$ is an
  appropriate non-hydrostatic vector added to ensure that the term in
  parenthesis has exactly one bare (without derivatives) $"d_{"scB}g_{"sM"sN}$
  in its divergence to match-up with the RHS of \cref{adiabaticity}. We can work
  out the corresponding constitutive relations, called Class H$_S$ for
  hydrostatic scalars, by noting that:
  \begin{equation}\label{varyN}
    \begin{split}
      \N_\sM ("cN "b^\sM)
      &= \frac{1}{\sqrt{-g}} "d_{"scB} (\sqrt{-g} "cN) \\
      &= \half \cN g^{\sM\sN} "d_{"scB} g_{\sM\sN} + \frac{\dow \cN}{\dow
        g_{\sM\sN}} "d_{"scB} g_{\sM\sN} + \frac{\dow \cN}{\dow (\dow_\sR
        g_{\sM\sN})} \dow_\sR "d_{"scB} g_{\sM\sN}
      + \ldots \\
      &\equiv \half T^{\sM\sN}_{"rmH_S} "d_{"scB}g_{\sM\sN} - \N_\sM
      "Q_{"cN}^\sM.
    \end{split}
  \end{equation}
  Here $\N_\sM "Q_{"cN}^\sM$ is a total derivative term gained after successive
  differentiation by parts, which defines $"Q_{"cN}^\sM$. Note that if $"cN$ has
  some total derivative terms, i.e. $"cN = \N_{"sM}X^{"sM}$ for an arbitrary
  hydrostatic vector $X^{"sM}$, upon choosing
  $"Q^{"sM}_{"cN} = - \frac{1}{\sqrt{-g}} \d_{"scB}(\sqrt{-g}X^{"sM})$ the free
  energy current
  $"cN "b^\sM + "Q_{"cN}^\sM = "b^{"sM}\N_{"sN}X^{"sN} - \frac{1}{\sqrt{-g}}
  \d_{"scB}(\sqrt{-g}X^{"sM}) = 2\N_{"sN}("b^{["sM}X^{"sN]})$ has zero
  divergence, and hence belongs to Class S. Class H$_S$ constitutive relations
  are therefore characterised by the most generic hydrostatic scalar $"cN$ up to
  total derivatives.

  On the other hand, $"bbN^\sM$ contains all the hydrostatic vectors transverse
  to $u^\sM$ and $V^\sM$ whose divergence contains exactly one bare
  $"d_{"scB}g_{"sM"sN}$. Intuitively, this means that we want to find
  ``spatial'' vectors without any ``time'' derivative in them, whose divergence
  however still contains a ``time'' derivative. Slight thought will reveal that
  this is only possible in parity-odd sector, which we are not considering in
  this work. For completeness, we should mention that they contain contributions
  from \emph{anomalies} (Class A) and some other parity-odd terms commonly
  dubbed as \emph{transcendental anomalies} (Class H$_V$)
  (see~\cite{Jain:2015jla} for more details), and are totally determined up to
  some constants.
  
\item \textbf{Non-hydrostatic transport (Classes D and $\overline{\text{D}}$):}
  These are solutions of \cref{adiabaticity} made purely out of non-hydrostatic
  tensor structures. They are best expressed by introducing a symmetric
  covariant derivative operator
  $"DD^n_{\sM_1\sM_2\ldots \sM_n} = \N_{(\sM_1}\N_{\sM_2}\ldots
  \N_{\sM_n)}$. They form a basis for arbitrary derivative operators because
  antisymmetric combinations can always be replaced by combinations of Riemann
  curvature tensor. We can now write down the most generic non-hydrostatic
  constitutive relations as combinations of $"DD^n "d_{"scB} g_{\sR\sS}$ for all
  $n$. A particularly convenient parametrisation is
  \begin{equation}\label{non-hydrostatic-consti}
    T^{\sM\sN}_{\text{non-hydrostatic}}
    =
    - \half \sum_{n=0}^\infty \lB
    "fC_n^{("sM"sN)("sR"sS)} \half "DD^n "d_{"scB} g_{\sR\sS}
    + "DD^n \lb "fC_n^{("sM"sN)("sR"sS)} \half "d_{"scB} g_{\sR\sS} \rb \rB,
  \end{equation}
  where $"fC_n^{("sM"sN)("sR"sS)}$ is an arbitrary matrix with $n$ additional
  symmetric indices to be contracted with $"DD^n$. Note that when $"DD^n$ in the
  second term hits $"d_{"scB} g_{\sR\sS}$, we get the same term as the first
  one. All the other terms coming out of differentiation by parts have merely
  been included for convenience, as we shall see. Some minor comments about the
  structure of $"fC_n^{("sM"sN)("sR"sS)}$: (1) It cannot contain an instance of
  $"DD^m g_{"sA"sB}$ for $m>n$ as the respective terms in
  \cref{non-hydrostatic-consti} are taken care of in the
  $"fC_m^{("sM"sN)("sR"sS)}$ term. (2) For an instance of $"DD^n g_{"sA"sB}$ in
  $"fC_n^{("sM"sN)("sR"sS)}$, we must be careful not to over-count the terms
  gained by $("sM"sN)\leftrightarrow("sA"sB)$ exchange, which would give the
  same contribution to $T^{\sM\sN}_{\text{non-hydrostatic}}$ in
  \cref{non-hydrostatic-consti}.

  We can make our lives much
  easier by thinking of $"fC_n^{("sM"sN)("sR"sS)}$ as a $\half(d+1)(d+2)$
  dimensional matrix, and $T^{\sM\sN}_{"rmD\cup\overline{"rmD}}$ and
  $"d_{"scB} g_{\sR\sS}$ as column vectors. By suppressing the indices in this
  notation, the above expression becomes
  \begin{equation}\label{non-hydrostatic-consti-matrix}
    T_{\text{non-hydrostatic}}
    =
    - \frac14 \sum_{n=0}^\infty \Big[
    "fC_n \cdot "DD^n "d_{"scB} g
    + "DD^n \lb "fC_n \cdot "d_{"scB} g \rb \Big].
  \end{equation}
  Using differentiation by
  parts on the second term we can compute
  \begin{align}\label{DbarDdecomposition}
    \half "d_{"scB} g^{"rmT} \cdot T_{\text{non-hydrostatic}}
    &=
      - \frac18 \sum_{n=0}^\infty \lB
      "d_{"scB} g^{"rmT} \cdot "fC_n \cdot "DD^n "d_{"scB} g
      + "d_{"scB} g^{"rmT} \cdot "DD^n \lb "fC_n\cdot "d_{"scB} g \rb \rB, \nn\\
    &= - \frac18
      "d_{"scB} g^{"rmT} \cdot \sum_{n=0}^\infty \lb "fC_n^{"rmT}
      + {(-)}^n "fC_n^{"rmT} \rb \cdot "DD^n "d_{"scB} g
      + \N_{"sM}"Q^{"sM}_{"fC}, \nn\\
    &= - \frac14
      "d_{"scB} g^{"rmT} \cdot \sum_{n=0}^\infty "fD_n \cdot "DD^n "d_{"scB} g
      + \N_{"sM}"Q^{"sM}_{"fD} + \N_{"sM}"Q^{"sM}_{\overline{"fD}}.
  \end{align}
  In the last line we have split $"fC_n$ into
  \begin{equation}
    \fD_n = \half\lb \fC_n + {(-)}^n\fC_n^{"rmT} \rb, \qquad
    \overline\fD_n = \half\lb \fC_n - {(-)}^n\fC_n^{"rmT} \rb,
  \end{equation}
  and the corresponding contribution to $"Q^{"sM}_{"fC}$ into $"Q^{"sM}_{"fD}$
  and $"Q^{"sM}_{\overline{"fD}}$. Note that for the constitutive relations
  coupled to $\overline\fD_n$, called Class $\overline\rmD$ for non-dissipative,
  the story pretty much ends here. We can infer from \cref{DbarDdecomposition}
  that the associated constitutive relations $T^{"sM"sN}_{\overline\rmD}$
  satisfy \cref{adiabaticity} with $N^{"sM} = "Q^{"sM}_{\overline{"fD}}$ and
  $"D = 0$. They correspond to non-hydrostatic transport that does not cause any
  dissipation.

  For the constitutive relations coupled to $"fD_n$ however, called Class D for
  dissipative, we need to do little more work to ensure that the associated $"D$
  can be made positive definite. To do that, we rewrite the $"fD_n$ part of
  \cref{DbarDdecomposition} as
  \begin{align}\label{firstDquadform}
    \half "d_{"scB} g^{"rmT} \cdot T_{"rmD}
    &= - \frac14 "d_{"scB} g^{"rmT} \cdot "fD_{0(0)} \cdot "d_{"scB} g
      - \frac12 "d_{"scB} g^{"rmT} \cdot
      "fD_{0(0)} \cdot \lb "U_1 \cdot "d_{"scB} g \rb 
      + \N_{"sM}"Q^{"sM}_{"fD} \nn\\
    &= - \underbrace{
      \frac14 \lb
      \lb 1 + "U_1 \rb \cdot "d_{"scB} g \rb^{"rmT} \cdot
      "fD_{0(0)} \cdot \lb \lb 1 + "U_1 \rb \cdot "d_{"scB} g \rb
    }_{\text{quadratic form}} \nn\\
    &\qquad\qquad + \underbrace{
      \frac14 \lb "U_1 \cdot "d_{"scB} g \rb^{"rmT} \cdot "fD_{0(0)} \cdot
      \lb "U_1 \cdot "d_{"scB} g \rb
    }_{\text{residue}} +
    \underbrace{\N_{"sM}"Q^{"sM}_{"fD}}_{\text{total derivative}},
  \end{align}
  where $"fD_{0(n)}$ denotes the $n$th derivative piece in $"fD_{(0)}$, while
  $"U_1$ is a differential operator
  \begin{equation}
    "U_1 = \half "fD_{0(0)}^{-1} \sum_{n=1}^\infty \lb "fD_{0(n)} + "fD_n "DD^n \rb.
  \end{equation}
  The quadratic form piece in \cref{firstDquadform} is of most interest to us,
  as it contributes to $"D$. The total derivative piece on the other hand is a
  contribution to the free energy current $N^{"sM}$. Finally, the residue piece
  is what we will like to get rid of. Note that every term in $"U_1$ has at
  least one derivative. Consequently, the residue piece is at least 4 order in
  derivatives. If we are only interested in the constitutive relations up to
  second derivative order, we can ignore this piece altogether. However, we will
  illustrate the full procedure here for completeness. Using differentiation by
  parts, the residue piece can be rewritten as
  \begin{equation}
    \frac14 \lb "U_1 \cdot "d_{"scB} g \rb^{"rmT} \cdot "fD_{0(0)} \cdot
    \lb "U_1 \cdot "d_{"scB} g \rb
    = \frac14 "d_{"scB} g^{"rmT} \cdot
    \lb "U_1^{\dagger} \cdot "fD_{0(0)} \cdot "U_1 \cdot "d_{"scB} g \rb
    + \N_{"sM}"Q^{"sM}_{"fD,1}.
  \end{equation}
  Putting this back in \cref{firstDquadform} we get
  \begin{align}\label{secondDquadform}
    \half "d_{"scB} g^{"rmT} \cdot T_{"rmD}
    &= - \underbrace{
      \frac14 \lb
      \lb 1 + "U_1 + "U_2 \rb \cdot "d_{"scB} g \rb^{"rmT} \cdot
      "fD_{0(0)} \cdot \lb \lb 1 + "U_1 + "U_2 \rb \cdot "d_{"scB} g \rb
      }_{\text{quadratic form}} \nn\\
    &\qquad
      + \underbrace{
      \frac12 \lb \lb "U_1 + \frac12 "U_2 \rb \cdot
      "d_{"scB} g \rb^{"rmT} \cdot
      "fD_{0(0)} \cdot \lb "U_2 \cdot "d_{"scB} g \rb
      }_{\text{residue}}
    +
      \underbrace{
      \N_{"sM} \lb "Q^{"sM}_{"fD} + "Q^{"sM}_{"fD,1} \rb
      }_{\text{total derivative}},
  \end{align}
  where $"U_2$ is another derivative operator
  \begin{equation}
    "U_2 = - \half "fD_{0(0)}^{-1} \cdot "U_1^{\dagger} \cdot "fD_{0(0)} \cdot "U_1.
  \end{equation}
  Comparing \cref{secondDquadform} to \cref{firstDquadform}, hopefully the
  reader can see a pattern. The quadratic form piece now has some additional
  higher derivative terms, whereas we have pushed the residue piece to 5th
  derivative order. We can repeat this procedure iteratively to push
  the residue piece to arbitrarily high derivative orders and obtain
  \begin{align}
    \half "d_{"scB} g^{"rmT} \cdot T_{"rmD}
    &= - \underbrace{
      \frac14 \lb
      \lb 1 + \sum_{n=1}^{\infty} "U_n \rb \cdot "d_{"scB} g \rb^{"rmT} \cdot
      "fD_{0(0)} \cdot \lb
      \lb 1 + \sum_{n=1}^{\infty} "U_n \rb \cdot "d_{"scB} g \rb
      }_{\text{quadratic form}} \nn\\
    &\qquad\qquad
    + \underbrace{
      \N_{"sM} \lb "Q^{"sM}_{"fD} + \sum_{n=1}^{\infty} "Q^{"sM}_{"fD,n} \rb
      }_{\text{total derivative}},
  \end{align}
  where
  \begin{equation}
    "U_{d+1} \big\vert_{d=1}^\infty
    = - \fD_{0(0)}^{-1} \cdot \lb \sum_{k=1}^{d-1} "U_k^\dagger
    + \half "U_d^\dagger  \rb \lb \fD_{0(0)} \cdot "U_d \rb.
  \end{equation}
  We see therefore that the Class D constitutive relations satisfy
  \cref{adiabaticity} with
  \begin{equation}
    "D = \frac14 \lb
    \lb 1 + \sum_{n=1}^{\infty} "U_n \rb \cdot "d_{"scB} g \rb^{"rmT} \cdot
    "fD_{0(0)} \cdot \lb
    \lb 1 + \sum_{n=1}^{\infty} "U_n \rb \cdot "d_{"scB} g \rb,
  \end{equation}
  and the free energy current 
  \begin{equation}
    N^{"sM} = "Q^{"sM}_{"fD} + \sum_{n=1}^{\infty} "Q^{"sM}_{"fD,n}.
  \end{equation}
  The condition $"D\geq 0$ therefore, only gives a constraint on the first
  derivative constitutive relations by forcing all the eigenvalues of
  $"fD_{0(0)}$ to be non-negative. See~\cite{Jain:2016rlz} for more details on
  this. We do not get any further constraints from the second law at higher
  derivative orders in the non-hydrostatic sector.
  
  Recall that to avoid the over-counting of constitutive relations related to
  each other by combinations of equations of motion, we had decided to drop all
  the constitutive relations that involve $u^{"sM}"d_{"scB}g_{"sM"sN}$. To
  respect this, we must demand that none of the $("sM"sN)("sR"sS)$ indices in
  $"fC^{("sM"sN)("sR"sS)}$ should come from a $u^{"sM}$. Equivalently
  $"fC^{("sM"sN)("sR"sS)} V_{"sM} = "fC^{("sM"sN)("sR"sS)} V_{"sR} =
  0$. Furthermore, $"fC^{("sM"sN)("sR"sS)}$ should of course not have an
  explicit occurrence of $u^{"sM}"d_{"scB}g_{"sM"sN}$.
\end{itemize}

\subsection{Results up to first order}\label{firsorderresults}

In~\cite{Banerjee:2015hra,Banerjee:2016qxf}, we discussed null fluids up to
first order in derivatives. We briefly recollect these results here and
illustrate how they fit into the classification presented in the previous
subsection. To setup the notation, we enlist the independent fluid data at first
order in \cref{first-order-data}.

\begin{table}[h]
    \renewcommand{\arraystretch}{1.6}
  \centering
  \begin{tabular}[t]{ccc}
    \midrule\midrule
    \multicolumn{3}{c}{Non-hydrostatic --- onshell independent} \\
    \midrule
    $"Q$
    & $\frac{T}2 P^{"sM"sN}"d_{"scB} g_{"sM"sN}$
    & $\N_{"sM} u^{"sM}$ \\
    
    $"t^{"sM}$, $\bar{"t}^{"sM}$
    & $TP^{"sM"sN}V^{"sR} "d_{"scB}g_{"sN"sR}$
    & $P^{"sM"sN}\frac{1}{T} \dow_{"sN} T$ \\
    
    $"s^{"sM"sN}$
    & $T P^{"sR\langle"sM}P^{"sN\rangle"sS} "d_{"scB}g_{"sR"sS}$
    & $2 P^{"sR\langle"sM}P^{"sN\rangle"sS} \N_{"sR} u_{"sS}$ \\
    
    \midrule\midrule
    \multicolumn{3}{c}{Non-hydrostatic --- onshell dependent} \\
    \midrule

    $"Q_{T}$
    & $T u^{"sM}V^{"sN}"d_{"scB} g_{"sM"sN}$
    & $u^{"sM} \frac1T \dow_{"sM} T$ \\
    
    $"Q_{"vp}$
    & $\frac{T}2 u^{"sM}u^{"sN}"d_{"scB} g_{"sM"sN}$
    & $u^{"sM} T \dow_{"sM} "vp$ \\

    $"t^{"sM}_{"vp}$
    & $TP^{"sM"sN}u^{"sR} "d_{"scB}g_{"sN"sR}$
    & $P^{"sM"sN}\lb T \dow_{"sN} "vp + u^{"sR}\N_{"sR}u_{"sN} \rb$ \\

    \midrule\midrule
    \multicolumn{3}{c}{Hydrostatic} \\
    
    \midrule

    $\bar{"t}^{"sM}_{"vp}$
    & \multicolumn{2}{c}{$P^{"sM"sN}T\dow_{"sN}"vp$} \\

    $"o^{"sM"sN}$
    & \multicolumn{2}{c}{$2 P^{"sR["sM}P^{"sN]"sS} \N_{"sR} u_{"sS}$} \\

    \midrule\midrule
  \end{tabular}
  \caption{\label{first-order-data} First order fluid data. Note that we have
    defined two symbols $"t^{"sM}$ and $\bar{"t}^{"sM}$ for the term
    $P^{"sM"sN}\frac{1}{T} \dow_{"sN} T$. This is to acknowledge the fact that
    in the presence of torsion, the term
    $\bar{"t}^{"sM} = P^{"sM"sN}\frac{1}{T} \dow_{"sN} T$ is actually
    hydrostatic, while
    $"t^{"sM} = P^{"sM"sN}\lb \frac{1}{T} \dow_{"sN} T + u^{"sR}H_{"sR"sN} \rb$
    is non-hydrostatic. This distinction will be handy later. We will also use
    the acceleration
    $a^{"sM} = u^{"sN}\N_{"sN}u^{"sM} = "t^{"sM}_{"vp} - \bar{"t}^{"sM}_{"vp}$
    sometimes.}
\end{table}

\begin{itemize}[leftmargin=1em]
\item \textbf{Hydrostatic transport (Class H$_{\text{S}}$):} The most generic
  hydrostatic scalar up to first derivative order is simply
  \begin{equation}
    \cN = P(T,"m_m),
  \end{equation}
  where $P(T,"m_m)$ is identified with the pressure of the fluid. Interestingly,
  there are no hydrostatic scalars involving just one derivative. We can use
  \cref{varyN} and find the associated constitutive relations, which basically
  gives the entire ideal null fluid
  \begin{equation}\label{firstOrder-HS}
    T^{"sM"sN}_{"rmH_S} = R u^{"sM}u^{"sN} + 2E u^{("sM}V^{"sN)} + P P^{"sM"sN}
    + "cO(\dow^2),
  \end{equation}
  with $"Q_{"sN}^{"sM} = 0$. Here $P^{"sM"sN} = g^{"sM"sN} + 2 u^{("sM}V^{"sN)}$
  is a projector transverse to $u^{"sM}$ and $V^{"sM}$.
  $R = \frac{1}{T}\frac{\dow P}{\dow "vp}$ is identified with the mass density
  and $E = T \frac{\dow P}{\dow T} - P$ with the energy density of the
  fluid. Together they define the thermodynamic relations
  \begin{equation}\label{thermodynamics}
    "dd E = T"dd S + "m_m "dd R, \qquad
    E + P = ST + R"m_m,
  \end{equation}
  where $S$ is called the entropy density.

\item \textbf{Non-hydrostatic transport (Classes D and $\overline{\text D}$):}
  Looking back at \cref{non-hydrostatic-consti}, we can infer that since
  $"d_{"scB}g_{"sM"sN}$ already contains a derivative, at first derivative order
  we only need a zero-derivative tensor $"fC^{("sM"sN)("sR"sS)}_{0}$. It takes
  the most generic form
  \begin{equation}\label{firstOrder-C0}
    "fC_0^{("sM"sN")("sR"sS)} =
    2T"h~P^{"sM\langle"sR} P^{"sS\rangle"sN}
    + T"z~P^{"sM"sN} P^{"sR"sS}
    + 4T^2"k~V^{("sM}P^{"sN)("sR}V^{"sS)}.
  \end{equation}
  First thing to note here is that all the terms in \cref{firstOrder-C0} are
  symmetric under the exchange of $("sM"sN)\leftrightarrow
  ("sR"sS)$. Consequently, the zero derivative $\overline{"fD}_0$ and hence the
  first derivative Class $\overline{"rmD}$ constitutive relations are
  identically zero,
  \begin{equation}\label{firstOrder-D}
    T^{"sM"sN}_{\overline{"rmD}} = "cO("pd^2), \qquad
    N^{"sM}_{\overline{"rmD}} = "cO("pd^2).
  \end{equation}
  On the other hand, for Class D we get
  \begin{equation}\label{firstOrder-Dbar}
    T^{"sM"sN}_{"rmD} = - "h "s^{"sM"sN} - "z P^{"sM"sN}"Q - 2T
    "k_{"e}V^{("sM} "t^{"sN)} + "cO("pd^2),
  \end{equation}
  with
  \begin{equation}
    N^{"sM}_{"rmD} = "cO("pd^2), \qquad
    "D = \frac{1}{2T} "h "s^{"sM"sN}"s_{"sM"sN}
    + \frac{1}{T} "z "Q^2 + "k_{"e} "t^{"sM} "t_{"sM}.
  \end{equation}
  The condition $"D\geq 0$ simply implies that all the transport coefficients are
  non-negative
  \begin{equation}\label{positivity-constraints}
    "h, \quad "z, \quad "k \quad \geq \quad 0.
  \end{equation}
  We can identify these transport coefficients as: $"h$ shear viscosity, $"z$
  bulk viscosity and $"k$ thermal conductivity of the fluid.
  
\end{itemize}

The full set of constitutive relations up to first derivative order are a
direct sum of \cref{firstOrder-HS,firstOrder-D,firstOrder-Dbar}, giving us
\begin{equation}
  T^{"sM"sN} = R u^{"sM}u^{"sN} + 2E u^{("sM}V^{"sN)} + P P^{"sM"sN}
  - "h "s^{"sM"sN} - "z P^{"sM"sN}"Q - 2T"k_{"e}V^{("sM}"t^{"sN)} + "cO("pd^2).
\end{equation}
They are supported by the free energy and entropy currents
\begin{equation}
  N^{"sM} = \frac{1}{T} P u^{"sM} + "cO("pd^2), \qquad
  J^{"sM}_S = N^{"sM} - T^{"sM"sN}"b_{"sN} = S u^{"sM} - "k_{"e}"t^{"sM} + "cO("pd^2).
\end{equation}
The transport coefficients follow the thermodynamic constraints in
\cref{thermodynamics} at ideal order and the positivity relations
\cref{positivity-constraints} at first derivative order.

This finishes our crash course in null fluids. We still need to discuss the
translation of these results to Galilean fluids, which we will come back to in
\cref{sec-redn}. In the next section, we will use the machinery developed here
to write down the null fluid constitutive relations up to second derivative
order.

\section{Second order null fluids}\label{sec:2ndNull}

In the previous section we gave a self contained review of null fluids and
presented an algorithm to generate the respective constitutive relations up to
arbitrarily high derivative orders. The goal of this section is to use this
algorithm to write down the null fluid constitutive relations up to second
order. Apart from $P$, $"h$, $"z$ and $"k$ at previous orders, we find a total
of 25 transport coefficients: 5 in Class H$_S$, 9 in Class D and 11 in Class
$\overline{\text D}$. For readers who are only interested in the results, they
have been summarised in
\cref{tab:2nd-order-hydrostatic,tab:2nd-order-hydrostatic-N,tab:1der-C0,tab:0der-C1,tab:0der-C1-N}.
In the remaining of this section, we will explain how we arrived at these
results.

\subsection{Hydrostatic transport (Class H$_{\text{S}}$)}

\begin{table}[!b]
  \renewcommand{\arraystretch}{2}
  \centering
  \begin{tabular}[h]{cc}
    \midrule\midrule
    \#
    & $\dsp T^{"sM"sN} \ni \frac{2}{\sqrt{-g}}\frac{"d(\sqrt{-g}"cN)}{"d g_{"sM"sN}}
      = "cN g^{"sM"sN} + 2\frac{"d"cN}{"d g_{"sM"sN}}$ \\
    \midrule\midrule

    $P$
    & $"fg^{"sM"sN}P \equiv R u^{"sM}u^{"sN} + 2E u^{("sM}V^{"sN)} + P P^{"sM"sN}$ \\
    
    \midrule
      
    $"a_1$
    & $\dsp \frac1{2T^2} \bar{"t}^{"sR}_{"vp} \bar{"t}_{"vp"sR}~"fg^{"sM"sN}("a_1 T^2)
      - \frac1T \N_{"sR}("a_1 T \bar{"t}^{"sR}_{"vp}) u^{"sM}u^{"sN}
      + 2"a_1 V^{("sM} \bar{"t}_{"vp}^{"sN)} "Q_{"vp}
      - "a_1 \bar{"t}_{"vp}^{("sM} \bar{"t}_{"vp}^{"sN)}$ \\[1ex]

    $"a_2$
    &
      \begin{minipage}[c]{0.8\textwidth}
        \centering
        $\dsp \bar{"t}^{"sR}_{"vp} \bar{"t}_{"sR} ~"fg^{"sM"sN}"a_2 - 2T
        \N_{"sR}\lb\frac{"a_2}{T} \bar{"t}^{"sR}_{"vp}\rb u^{("sM}V^{"sN)} - \frac1T
        \N_{"sR}("a_2T \bar{"t}^{"sR}) u^{"sM}u^{"sN}$ \\
        $\dsp + 2"a_2 V^{("sM} \bar{"t}_{"vp}^{"sN)} "Q_{T}
        + 2"a_2 V^{("sM} \bar{"t}^{"sN)} "Q_{"vp}
        - 2"a_2 \bar{"t}^{("sM} \bar{"t}_{"vp}^{"sN)}$
      \end{minipage} \\[3ex]

    $"a_3$
    & $\dsp\frac14 "o^{"sR"sS}"o_{"sR"sS} "fg^{"sM"sN}"a_3
      - "a_3 "o^{"sM}_{~"sR} "o^{"sN"sR}
      + 2"a_3 V^{("sM}"o^{"sN)"sR} a_{"sR}
      - 2 \N_{"sR}\lb "a_3 "o^{"sR"sS} \rb P_{"sS}{}^{("sM} u^{"sN)}$ \\
    
    $"a_4$
    & $\dsp - 2 \N_{"sR}\lb "a_4"o^{"sR"sS} \rb P_{"sS}{}^{("sM} V^{"sN)}$ \\

    $"a_5$
    & $\dsp R"fg^{"sM"sN}"a_5 - 2R^{"sM"sN}"a_5
      + 2 \N^{"sM}\N^{"sN}"a_5 - 2 g^{"sM"sN}\N^{"sR}\N_{"sR}"a_5$ \\

    \midrule\midrule
    
  \end{tabular}
  \caption{\label{tab:2nd-order-hydrostatic} Class H$_S$ constitutive relations
    up to second derivative order. We have defined a differential operator
    $"fg^{"sM"sN} = g^{"sM"sN} + 2T u^{("sM}V^{"sN)} \frac{"pd}{"pd T} +
    \frac{1}{T} u^{"sM}u^{"sN} \frac{"pd}{"pd "vp}$ for brevity.}

  \vspace{2ex}
  \begin{tabular}[t]{cc}
    \midrule\midrule
    \#
    & $\dsp N^{"sM} \ni "cN "b^{"sM} + "Q_{"cN}^{"sM} $ \\
    \midrule\midrule

    $P$
    & $\dsp \frac{1}{T} P u^{"sM}$ \\
    
    \midrule

    $"a_1$
    & $\dsp \frac{"a_1}T\lb \half u^{"sM} ~\bar{"t}^{"sR}_{"vp} \bar{"t}_{"vp"sR}
      - \bar{"t}^{"sM}_{"vp} "Q_{"vp} \rb$ \\

    $"a_2$
    & $\dsp \frac{"a_2}T\lb u^{"sM} \bar{"t}^{"sR}_{"vp} \bar{"t}_{"sR}
      - \bar{"t}^{"sM} "Q_{"vp} - \bar{"t}^{"sM}_{"vp} "Q_{T} \rb$ \\

    $"a_3$
    & $\dsp \frac{"a_3}T \lB \frac14 u^{"sM} "o^{"sR"sS}"o_{"sR"sS}
      - "o^{"sM"sN} "t_{"vp"sN} \rB$ \\

    $"a_4$
    & $\dsp - \frac{"a_4}T "o^{"sM"sN}"t_{"sN}$ \\

    $"a_5$
    & $\frac{"a_5}{T} R u^{"sM}
      - 2"a_5 \N_{"sR}\N^{("sR}\bfrac{u^{"sM)}}{T}
      + 2\N^{("sR}\bfrac{u^{"sM)}}{T} \dow_{"sR}"a_5
      + 2"a_5\N^{"sM} \N_{"sR}\bfrac{u^{"sR}}{T}
      - 2 \N_{"sR}\bfrac{u^{"sR}}{T} \N^{"sM}"a_5$ \\

    \midrule\midrule
    
  \end{tabular}
  \caption{\label{tab:2nd-order-hydrostatic-N} Class H$_S$ free energy current
    up to second derivative order.}
\end{table}

Let us start with Class H$_S$. As discussed in \cref{sec:classification}, this
class is characterised by the most generic hydrostatic scalar $"cN$ up to some
total derivative terms. Up to second derivative order, we have 5 such terms
apart from the ideal order pressure
\begin{equation}\label{second-order-N}
  \cN
  =
  P
  + \frac12 "a_1 \bar{"t}^{"sM}_{"vp} \bar{"t}_{"vp"sM}
  + "a_2 \bar{"t}^{"sM}_{"vp} "t_{"sM}
  + \frac14 "a_3 "o^{"sM"sN}"o_{"sM"sN}
  + \half "a_4 "o^{\sM\sN} H_{"sM"sN}
  + "a_5 R + "cO(\dow^3),
\end{equation}
where $R$ is the Ricci scalar. Note the second and forth terms in particular:
the former is non-hydrostatic while the latter vanishes. However, their
variations do contribute to hydrostatic constitutive relations as we shall see
in the following. This is an instance of the ``unnaturalness'' arising due to
the absence of torsion we talked about around \cref{RiemannDef}. In the presence
of torsion, both of these terms are actually hydrostatic. Astute reader might
note that by this token we should also include the quadratic terms coupling to
$"t^{"sM}"t_{"sM}$ and $H^{"sM"sN}H_{"sM"sN}$. However, in the absence of
torsion their variations only lead to non-hydrostatic constitutive relations and
thus can be ignored here.

Varying $"cN$ in \cref{second-order-N} and using the formulas in \cref{varyN},
we can now read out the Class H$_S$ constitutive relations and free energy
current. Note that the $"d_{"scB}$ variation of fluid variables can be
represented in terms of $"d_{"scB} g_{"sM"sN}$ as
\begin{align}
  "d_{"scB} T
  &= T V^\sM u^\sN "d_{"scB} g_{\sM\sN},\quad
    "d_{"scB} \varpi = \frac{1}{2T} u^\sM u^\sN "d_{"scB} g_{\sM\sN}, \quad
    "d_{"scB} u^{\sM} = \lb 2u^{\sM} V^{(\sR} u^{\sS)}
    + V^{"sM} u^{"sR}u^{"sS} \rb \half "d_{"scB} g_{\sR\sS}, \nn\\
  "d_{"scB} u_{"sM}
  &= \lb - V_{"sM} u^{"sR} u^{"sS}
    + 2 P_{"sM}^{~("sR} u^{"sS)} \rb \half "d_{"scB}g_{"sR"sS}, \quad
    "d_{"scB}V_{"sM} = \lb - 2 V_{"sM} u^{("sR} V^{"sS)}
    + 2 P_{"sM}^{~("sR} V^{"sS)} \rb \half "d_{"scB} g_{"sR"sS},
\end{align}
while that of Ricci scalar as
\begin{equation}
  \d_{"scB} R =
  \lb -2R^{"sM"sN} + 2\N^{("sM} \N^{"sN)}
  - 2g^{"sM"sN} \N^{"sR}\N_{"sR} \rb \half "d_{"scB} g_{"sR"sS}.
\end{equation}
These will be useful in the variational calculation. We have summarised our
results in \cref{tab:2nd-order-hydrostatic,tab:2nd-order-hydrostatic-N}. They
enlist the term by term contribution to $T^{"sM"sN}$ and $N^{"sM}$ coming from
varying \cref{second-order-N}. The full Class H$_{S}$ constitutive relations
and free energy current will be a direct sum of all these contributions.

\subsection{Non-hydrostatic transport (Classes D and $\overline{\text D}$)}

Up to second derivative order, the non-hydrostatic constitutive relations in
\cref{non-hydrostatic-consti} get contributions from a zero derivative tensor
$"fC_1^{("sM"sN)("sR"sS)"sT}$ and a tensor
$"fC_0^{("sM"sN)("sR"sS)}$ with at most one derivative
\begin{equation}
  T^{\sM\sN}_{"rmD\cup\overline{"rmD}}
  =
  - "fC_0^{("sM"sN)("sR"sS)} \half "d_{"scB} g_{\sR\sS}
  - "fC_ 1^{("sM"sN)("sR"sS)"sT} \half \N_{"sT} "d_{"scB} g_{\sR\sS} 
  - \half \N_{"sT} "fC_1^{("sM"sN)("sR"sS)"sT} \half "d_{"scB} g_{\sR\sS} + "cO(\dow^3).
\end{equation}
The respective contribution to the free energy current truncated to two
derivatives is given as
\begin{equation}
  N^{"sM}_{"rmD\cup\overline{"rmD}} = - \frac18 "fC_{1}^{("sR"sS)("sA"sB)"sM}
  "d_{"scB} g_{\sR\sS} "d_{"scB} g_{\sA\sB}  + "cO(\dow^3).
\end{equation}
Interestingly, it is only non-vanishing for the Class $\overline{\text D}$
constitutive relations associated with $\overline{"fD}_{1}^{("sR"sS)("sA"sB)"sM}
= "fC_{1}^{("sR"sS)("sA"sB)"sM}$. Class D constitutive relations up to second
derivative order do not need any free-energy transport.

\begin{table}[!t]
  \renewcommand{\arraystretch}{1.6}
  \centering
  \begin{tabular}[h]{cccc}
    \midrule\midrule
    \#
    & \multicolumn{2}{c}{$"fC_0^{("sM"sN")("sR"sS)}$}
    & $T^{"sM"sN} \ni - "fC_0^{("sM"sN")("sR"sS)} \half "d_{"scB} g_{"sR"sS} $ \\
    \midrule\midrule

    $"h$
    & \multicolumn{2}{c}{$2T"h~P^{"sM\langle"sR} P^{"sS\rangle"sN}$}
    & $- "h "s^{"sM"sN}$ \\
    $"z$
    & \multicolumn{2}{c}{$T"z~P^{"sM"sN} P^{"sR"sS}$}
    & $- "z~P^{"sM"sN} "Q$ \\
    $"k T$
    & \multicolumn{2}{c}{$4T^2 "k~V^{("sM}P^{"sN)("sR}V^{"sS)}$}
    & $- 2T "k~V^{("sM} "t^{"sN)}$ \\

    \midrule
    
    $"d^+_1$
    & $2T^2"d^+_1~\lb V^{(\sM}P^{\sN)\langle"sR} P^{"sS\rangle("sA} V^{"sB)}\rb "d_{"scB} g_{"sA"sB}$
    & $2T"d^+_1~V^{(\sM}P^{\sN)\langle"sR}"t^{"sS\rangle}$
    & $- "d^+_1~V^{(\sM} "s^{"sN)"sR} "t_{"sR}$ \\
    
    $"d^-_1$
    & $2T^2 "d^-_1~\lb V^{("sR} P^{"sS)\langle"sM} P^{"sN\rangle("sA} V^{"sB)}
    \rb "d_{"scB} g_{"sA"sB}$
    & $2T "d^-_1~"t^{\langle"sM} P^{"sN\rangle ("sR} V^{"sS)}$
    & $- "d^-_1~"t^{\langle"sM} "t^{"sN\rangle}$ \\
    
    $"d^+_2$
    & $2T^2 "d^+_2~\lb V^{(\sM}P^{\sN)("sA}V^{"sB)} P^{\sR\sS} \rb "d_{"scB}g_{"sA"sB}$
    & $2T "d^+_2~V^{(\sM} "t^{\sN)} P^{\sR\sS}$
    & $- 2"d^+_2~V^{(\sM} "t^{\sN)} "Q$ \\

    $"d^-_2$
    & $2T^2 "d^-_2~\lb V^{(\sR} P^{\sS)("sA} V^{"sB)} P^{\sM\sN} \rb "d_{"scB} g_{"sA"sB}$
    & $2T "d^-_2~P^{\sM\sN} V^{(\sR} "t^{\sS)}$ 
    & $- "d^-_2~P^{\sM\sN} "t^{"sR}"t_{"sR}$ \\

    $"d^+_3$
    & $2T^2 "d^+_3~\lb P^{"sM\langle "sA}P^{"sB\rangle "sN}P^{"sR"sS}\rb "d_{"scB}g_{"sA"sB}$
    & $2T "d^+_3~"s^{"sM"sN} P^{"sR"sS}$
    & $- 2"d^+_3~"s^{"sM"sN} "Q$ \\

    $"d^-_3$
    & $ 2T^2 "d^-_3~\lb P^{"sR\langle"sA}P^{"sB\rangle"sS}P^{"sM"sN}\rb "d_{"scB}g_{"sA"sB}$
    & $2T "d^-_3~P^{"sM"sN} "s^{"sR"sS}$
    & $- "d^-_3~P^{"sM"sN} "s^{"sR"sS}"s_{"sR"sS} $ \\

    $"d_4$
    & $ 2T^2 "d_4~\lb P^{"sA\langle"sM}P^{"sB\langle "sR}P^{"sS\rangle"sN\rangle} \rb
    "d_{"scB} g_{"sA"sB}$
    & $2 T "d_4~"s^{\langle"sM\langle"sR} P^{"sS\rangle"sN\rangle}$
    & $- "d_4~"s^{\langle"sM"sR} "s_{"sR}^{\ "sN\rangle}$ \\

    $"d_5$
    & $T^2 "d_5~\lb P^{"sM"sN} P^{"sR"sS} P^{"sA"sB} \rb "d_{"scB} g_{"sA"sB}$
    & $2T "d_5~P^{"sM"sN} P^{"sR"sS} "Q$
    & $- 2 "d_5~P^{"sM"sN} "Q^2$ \\
    
    \midrule

    $"d^+_6$
    & \multicolumn{2}{c}
    {$2T "d^+_6~V^{(\sM}P^{\sN)\langle"sR} \bar{"t}^{"sS\rangle}_{"vp}$}
    & $- "d^+_6~V^{(\sM} "s^{"sN)"sR} \bar{"t}_{"vp"sR} $ \\

    $"d^-_6$
    & \multicolumn{2}{c}
    {$2T^2 "d^-_6~ \bar{"t}_{"vp}^{\langle"sM} P^{"sN\rangle ("sR} V^{"sS)}$}
    & $- "d^-_6~\bar{"t}^{\langle"sM}_{"vp} "t^{"sN\rangle}$ \\
    
    $"d^+_7$
    & \multicolumn{2}{c}
    {$2T^2 "d^+_7~V^{(\sM} \bar{"t}_{"vp}^{\sN)} P^{\sR\sS}$}
    & $- 2 "d^+_7~V^{(\sM} \bar{"t}_{"vp}^{\sN)} "Q $ \\
    
    $"d^-_7$
    & \multicolumn{2}{c}
    {$2T^2 "d^-_7~P^{\sM\sN} V^{(\sR} \bar{"t}_{"vp}^{\sS)}$}
    & $- "d^-_7~P^{\sM\sN} \bar{"t}^{"sR}_{"vp} "t_{"sR}$ \\
    
    $\bar{"d}_8$
    & \multicolumn{2}{c}{$2T \bar{"d}_8~V^{("sM} "o^{"sN)("sR}V^{"sS)}$}
    & $- \bar{"d}_8~V^{("sM} "o^{"sN)"sR} "t_{"sR}$ \\
    
    $\bar{"d}_9$
    & \multicolumn{2}{c}{$2T\bar{"d}_9~"o^{\langle "sM\langle "sR} P^{"sS\rangle "sN \rangle}$} 
    & $- \bar{"d}_9~"o^{\langle "sM "sR} "s_{"sR}{}^{"sN \rangle}$ \\
    \midrule\midrule
  \end{tabular}
  \caption{\label{tab:1der-C0} Terms in $"fC_0^{("sM"sN")("sR"sS)}$ with at most
    one derivative and their contribution to the second order Class D and
    $\overline{\text D}$ constitutive relations. Here we have defined
    $"d^{\pm}_i = ("d_i \pm \bar{"d}_i)/2$. The transport coefficients $"d_i$ couple
    to Class D constitutive relations while $\bar{"d}_i$ couples to Class
    $\overline{\text D}$.}
\end{table}

\begin{table}[!p]
  \renewcommand{\arraystretch}{2}
  \centering
  
  \begin{tabular}[h]{ccc}
    \midrule\midrule
    \# &
    $"fC_1^{("sM"sN)("sR"sS)"sT}$ 
    & $T^{"sM"sN} \ni
      - "fC_1^{("sM"sN)("sR"sS)"sT} \half \N_{"sT} "d_{"scB} g_{"sR"sS}
      - \half \N_{"sT} "fC_1^{("sM"sN)("sR"sS)"sT} \half "d_{"scB} g_{"sR"sS}$ \\[1ex]

    \midrule\midrule

    $"d^+_{10}$
    & $2T"d^+_{10}~P^{"sT("sM} V^{"sN)} P^{"sR"sS}$
    & $\dsp - 2V^{("sM}P^{"sN)"sR} \lB
      "d^+_{10}\dow_{"sR}"Q + \frac{T}2 "Q \dow_{"sR}\bfrac{"d^+_{10}}{T}\rB
      + "d^+_{10} V^{("sM} \N^{"sN)} u_{"sR} "t^{"sR}$ \\[2ex]

    $"d^-_{10}$
    & $- 2T"d^-_{10}~P^{"sM"sN} V^{("sR} P^{"sS)"sT}$
    & $\dsp P^{"sM"sN} \lB "d^-_{10} \N_{"sR}"t^{"sR}
      + \frac{T}2 "t^{"sR}\dow_{"sR} \bfrac{"d^-_{10}}{T} \rB
      + "d^-_{10} V^{("sM} \N_{"sR} u^{"sN)} "t^{"sR}$ \\[2ex]

    $"d^+_{11}$
    & $2T"d^+_{11}~V^{("sM} P^{"sN)\langle "sR} P^{"sS\rangle "sT}$
    &
      \begin{minipage}[c]{0.6\textwidth}
        \centering
        $\dsp - V^{("sM} P^{"sN)}{}_{"sR} \lB
        "d^+_{11}\N_{"sT} "s^{"sR"sT}
        + \frac{T}2 "s^{"sR"sT} \dow_{"sT}\bfrac{"d^+_{11}}{T} \rB$ \\
        $\dsp + \frac12 "d^+_{11} V^{("sM} \lb
        \N_{"sR} u^{"sN)}
        + P^{"sN)}{}_{"sR} "Q
        - \frac{2}{d}\N^{"sN)} u_{"sR}
        \rb "t^{"sR} $
      \end{minipage} \\[5ex]

    $"d^-_{11}$
    & $- 2T "d^-_{11}~P^{"sT\langle"sM} P^{"sN\rangle("sR} V^{"sS)}$
    &
      \begin{minipage}[c]{0.6\textwidth}
        \centering
        $\dsp P^{"sT\langle"sM}P^{"sN\rangle}{}_{"sR} \lB
        "d^-_{11} \N_{"sT}"t^{"sR}
        + \frac{T}2 "t^{"sR}\dow_{"sT}\bfrac{"d^-_{11}}{T} \rB$\\
        $\dsp + \frac12 "d^-_{11} V^{("sM} \lb \N^{"sN)} u_{"sR}
        + P^{"sN)}{}_{"sR} "Q 
        - \frac{2}{d} \N_{"sR} u^{"sN)} \rb "t^{"sR}$
      \end{minipage} \\[5ex]

    $\bar{"d}_{12}$
    & $2T \bar{"d}_{12}~P^{"sM"sN} P^{"sR"sS} u^{"sT}$
    &
      \begin{minipage}[c]{0.6\textwidth}
        \centering
        $\dsp - 2P^{"sM"sN} \lB \bar{"d}_{12} u^{"sR}\dow_{"sR}"Q + \frac{T}{2}
        "Q u^{"sR} \dow_{"sR}\bfrac{\bar{"d}_{12}}{T} \rB$\\
        $\dsp
        - P^{"sM"sN} \bar{"d}_{12}\lb "Q^2 - a^{"sR}"t_{"sR}\rb
        - 2\bar{"d}_{12} V^{("sM} a^{"sN)} "Q$
      \end{minipage} \\[5ex]

    $\bar{"d}_{13}$
    & $2T \bar{"d}_{13}~P^{"sM\langle"sR} P^{"sS"\rangle"sN} u^{"sT}$
    & \begin{minipage}[c]{0.6\textwidth}
      \centering
      $\dsp - P^{"sM}{}_{"sR}P^{"sN}{}_{"sS}\lB
      \bar{"d}_{13} u^{"sR}\N_{"sR} "s^{"sR"sS}
      + \frac{T}{2} "s^{"sR"sS}
      u^{"sT}\dow_{"sT}\bfrac{\bar{"d}_{13}}{T} \rB$
      $\dsp
      - \bar{"d}_{13}
      \lb\frac{1}{2} "s^{"sM"sN}"Q - a^{\langle"sM}"t^{"sN\rangle}\rb
      - \bar{"d}_{13} V^{("sM} "s^{"sN)"sR}a_{"sR}$
      \end{minipage} \\[3ex]

    \midrule\midrule
  \end{tabular}
  \caption{\label{tab:0der-C1} Zero derivative terms in
    $"fC_1^{("sM"sN")("sR"sS)"sT}$ and their contribution to the second order
    Class D and $\overline{\text D}$ constitutive relations. Here we have
    defined $"d^{\pm}_i = ("d_i \pm \bar{"d}_i)/2$. The transport coefficients
    $"d_i$ couple to Class D constitutive relations while $\bar{"d}_i$ couples
    to Class $\overline{\text D}$.}

  \vspace{5ex}
  \begin{tabular}[h]{cc}
    
    \midrule\midrule
    \#
    & $\dsp N^{"sM} \in - \frac18 "fC_{1}^{("sR"sS)("sA"sB)"sM}
      "d_{"scB} g_{\sR\sS} "d_{"scB} g_{\sA\sB}$ \\
    \midrule\midrule
    
    $\bar{"d}_{10}$
    & $\dsp - \frac{\bar{"d}_{10}}{2T} "t^{"sM}"Q$ \\

    $\bar{"d}_{11}$
    & $\dsp - \frac{\bar{"d}_{11}}{4T} "s^{"sM"sN} "t_{"sN}$ \\

    $\bar{"d}_{12}$
    & $\dsp - \frac{\bar{"d}_{12}}{T} u^{"sM} "Q^2$ \\

    $\bar{"d}_{13}$
    & $\dsp - \frac{\bar{"d}_{13}}{4T} u^{"sM} "s^{"sM"sN}"s_{"sM"sN}$ \\
    
    \midrule\midrule
  \end{tabular}
  \caption{\label{tab:0der-C1-N} Class $\overline{\text D}$ free energy current
    up to second derivative order. This is the only contribution to the
    non-hydrostatic free energy current.}
  
\end{table}

Let us start our discussion with $"fC_0$. We need the most generic 4-tensor
$"fC_0^{("sM"sN)("sR"sS)}$ made out of $0$ and $1$-derivative data with appropriate
symmetry properties. Note that none of the indices can come from a $u^{"sM}$
because the respective terms are eliminated by the equations of motion. Note
also that there are $1$-derivative terms in $"fC_0$ which are constructed using
$"d_{"scB} g_{"sM"sN}$, i.e.
$\fC_{0}^{(\sM\sN)(\sR\sS)(\sA\sB)}"d_{"scB} g_{\sA\sB} \in
\fC_{0}^{(\sM\sN)(\sR\sS)}$. The respective contribution to $T^{"sM"sN}$ would
be
\begin{equation}
  \frac{1}{2} \fC_{0}^{(\sM\sN)(\sR\sS)(\sA\sB)}
  "d_{"scB} g_{\sA\sB} "d_{"scB} g_{\sR\sS}.
\end{equation}
However, we would get the same contribution to $T^{"sM"sN}$ if we started with a
$("sR"sS)\leftrightarrow("sA"sB)$ swapped term
$\fC_{0}^{(\sM\sN)(\sA\sB)(\sR\sS)}"d_{"scB} g_{\sA\sB} \in
\fC_{0}^{(\sM\sN)(\sR\sS)}$ instead. Therefore to avoid over-counting, we need
to ensure that if we include a term
$\fC_{0}^{(\sM\sN)(\sR\sS)(\sA\sB)}"d_{"scB} g_{\sA\sB}$ in
$\fC_{0}^{(\sM\sN)(\sR\sS)}$, we should drop out a corresponding term
$\fC_{0}^{(\sM\sN)(\sA\sB)(\sR\sS)}"d_{"scB} g_{\sA\sB}$ which would give the
same contribution to the constitutive relations. Keeping in mind this minor
technicality, we have enlisted all the possible terms in $"fC_0$ and their
contribution to $T^{"sM"sN}$ in \cref{tab:1der-C0}. There is no associated free
energy current.

Let us now move on to $"fC_1$. We need to write down the most generic 5-tensor
$"fC_ 1^{("sM"sN)("sR"sS)"sT}$ made out of $0$-derivative data. Note that the
index $"sT$ cannot come from a $V^{"sT}$ because the respective term will have a
contraction with $\N_{"sT}$ causing it to vanish. The remaining 4 indices cannot
come from a $u^{"sM}$ as the respective terms have been eliminated using the
equations of motion. The resultant allowed terms in $"fC_1$ have been enlisted
in \cref{tab:0der-C1} along with their contribution to the constitutive
relations. The respective contribution to the free energy current has been given
in \cref{tab:0der-C1-N}.

It should be noted that there is a plausible term
$"a V^{(\sM}P^{\sN)(\sR}V^{\sS)} u^{"sT}$ in $"fC_1$ which we have not included
in \cref{tab:0der-C1}. Its contribution to $T^{"sM"sN}$, after some
simplification, would have been
\begin{equation}
  - \half V^{("sM} \lB
  "a \N^{"sN)}\lb \frac{1}{T} "Q_{T}\rb
  - \frac{1}{T^2} "a \N^{"sN)} u^{"sR} \N_{"sR} T
  + \frac1{2T} "a "t^{"sN)} "Q
  + \frac1{2T} "t^{"sN)} u^{"sR}\dow_{"sR} "a \rB.
\end{equation}
Note that the last three terms are composites and are linearly dependent on the
contributions from $"fC_0$. The first term is pure derivative, but is made
linearly dependent by the equations of motion. Hence the contribution of
$"a V^{(\sM}P^{\sN)(\sR}V^{\sS)} u^{"sT}$ is not linearly independent
onshell. It is worth pointing out that this apparent exception is another
``unnatural'' consequence of working with no torsion. In presence of torsion,
the respective contribution would have included an independent term coupling to
$V^{("sM}u^{"sR}\N_{"sR}"t^{"sN)}$.

\subsection{Mass Frame}\label{sec:massframe}

In our analysis above, we had fixed the hydrodynamic redefinition freedom in
$u^{"sM}$, $T$ and $"m_m$ by relating them to $"b^{"sM}$, as defined in
\cref{offshell-hydro-frame}. Although this is a convenient choice of frame for
computations, it is not the most physically interesting one. A better physically
motivated choice is the so called ``mass frame'', in which the mass current
(after null reduction) does not get any derivative corrections. It is defined as
\begin{equation}\label{mass-frame-condition}
  T^{"sM"sN}_{\text{mf}} V_{"sN} = - R u^{"sM} - E V^{"sM}.
\end{equation}
Let us start with the results in our frame and schematically represent them as
\begin{equation}
  T^{"sM"sN} = R u^{"sM}u^{"sN} + 2E u^{("sM}V^{"sN)} + P P^{"sM"sN}
  + "cT^{"sM"sN} + "cO("dow^3),
\end{equation}
where the tensor $"cT^{"sM"sN}$ contains all the derivative corrections. A generic
change of frame would amount to a field redefinition
\begin{equation}
  u^{"sM} \to u^{"sM} + "d u^{"sM}, \qquad
  T \to u^{"sM} + "d T, \qquad
  "vp \to u^{"sM} + "d "vp.
\end{equation}
We can check that the first order null fluid automatically respects the mass
frame, so we only need to perform this redefinition for second order variations
$"d u^{"sM}$, $"d T$ and $"d "vp$. This immediately implies that $"cT^{"sM"sN}$
remains unchanged up to three derivative terms. Under a second order
redefinition therefore, the energy momentum tensor changes as
\begin{multline}\label{mass-frame-T-var}
  T_{\text{mf}}^{"sM"sN} =
  R u^{"sM}u^{"sN}
  + 2E u^{("sM}V^{"sN)} + P P^{"sM"sN}
  + 2R u^{("sM} "d u^{"sN)}
  + \lb \frac{\dow R}{\dow T} "d T + \frac{\dow R}{\dow "vp} "d "vp \rb
  u^{"sM}u^{"sN}
  + 2(E+P) V^{("sM} "d u^{"sN)} \\
  + 2\lb \frac{\dow E}{\dow T} "d T + \frac{\dow E}{\dow "vp} "d "vp \rb
  u^{("sM}V^{"sN)}
  + \lb \frac{E+P}{T} "d T + TR "d "vp \rb P^{"sM"sN}
  + "cT^{"sM"sN} + "cO("dow^3).
\end{multline}
The condition \bref{mass-frame-condition} then requires
\begin{equation}
  \begin{pmatrix}
    \frac{1}{T} "d T \\
    T "d "vp
  \end{pmatrix}
  =
  -
  \begin{pmatrix}
    T\frac{\dow E}{\dow T} & \frac{1}{T} \frac{\dow E}{\dow "vp} \\
    T\frac{\dow R}{\dow T} & \frac{1}{T} \frac{\dow R}{\dow "vp}
  \end{pmatrix}^{-1}
  \begin{pmatrix}
    "cT^{"sM"sN}u_{"sM}V_{"sN} \\
    "cT^{"sM"sN}V_{"sM}V_{"sN}
  \end{pmatrix}, \qquad
  "d u^{"sM} = \frac{1}{R} P^{"sM}{}_{"sN}"cT^{"sN"sR}V_{"sR}.
\end{equation}
After plugging these back into \cref{mass-frame-T-var}, we can get $T^{"sM"sN}$
in the mass frame
\begin{equation}
  T_{\text{mf}}^{"sM"sN} =
  R u^{"sM}u^{"sN}
  + 2E u^{("sM}V^{"sN)} + P P^{"sM"sN}
  + "cT^{"sM"sN}_{\text{mf}}
  + "cO(\dow^3),
\end{equation}
where
\begin{equation}\label{offshellToMassFrame}
  "cT^{"sM"sN}_{\text{mf}} = 2 V^{("sM} P^{"sN)}{}_{"sR} "cT^{"sR"sS} \lb
  \frac{E+P}{R} V_{"sS} - u_{"sS} \rb
  - "cT^{"sR"sS} \lb \frac{\dow P}{\dow E} u_{"sR}V_{"sS}
  + \frac{\dow P}{\dow R} V_{"sR}V_{"sS}\rb P^{"sM"sN}
  + P^{"sM}{}_{"sR} P^{"sN}{}_{"sS} "cT^{"sM"sN}.
\end{equation}

This finishes our discussion of second order null fluids. The respective
constitutive relations and free energy current, in offshell hydrodynamic frame,
have been summarised in
\cref{tab:2nd-order-hydrostatic,tab:2nd-order-hydrostatic-N,tab:1der-C0,tab:0der-C1,tab:0der-C1-N}. If
the reader is instead interested in the results in mass frame, the formula for
translation is given in \cref{offshellToMassFrame}. In the next section, we will
perform null reduction on these results to obtain the constitutive relations of
a second order Galilean fluid.


\section{Second order Galilean fluids via null reduction}\label{sec-redn}

We will now reduce our null fluid results presented in \cref{sec:2ndNull}, to
obtain the constitutive relations of a Galilean fluid up to second order. We
will mainly focus on the covariant Newton-Cartan notation to deal with Galilean
fluids coupled to curved backgrounds. Later, specialising to flat backgrounds,
we will also discuss the conversion of these results to the conventional
non-covariant notation. For more details, please refer to our previous work
\cite{Banerjee:2015hra}.

\subsection{Newton-Cartan backgrounds}

Newton-Cartan geometries are a covariant representation of spacetimes which
respect Galilean symmetries.  As we established in~\cite{Banerjee:2015hra},
Newton-Cartan backgrounds are related to null backgrounds by a mere choice of
basis. In the following we briefly review the argument, and in the process
introduce the reader to the basics of Newton-Cartan backgrounds. For a fuller
and excellent review of Newton-Cartan geometries, please refer the appendix
of~\cite{Jensen:2014aia}.

On a null background, we choose a basis $\{x^\sM\} = \{x^-,x^\mu\}$ such that
the null isometry $"scV = \{ V = \dow_- \}$. The fact that $"scV$ is an isometry
implies that all the fields in the theory are independent of the $x^-$
coordinate. To perform the reduction, we require an arbitrary null field $v^\sM$
normalized as $v^\sM v_\sM = 0$, $v^\sM V_\sM = -1$, which can be interpreted as
providing a ``Galilean frame of reference''. In the case of a null fluid, the
null fluid velocity $v^\sM = u^\sM$ defines a special Galilean frame which we
refer to as the ``fluid frame of reference''. In an arbitrary Galilean frame, we
decompose the fields $V^\sM$, $v^\sM$ and $g_{\sM\sN}$ in the chosen basis as
\begin{equation}
  V^\sM = \begin{pmatrix} 1 \\ 0 \end{pmatrix}, \qquad
  v^\sM = \begin{pmatrix} v^\mu B^{(v)}_\mu \\ v^\mu \end{pmatrix}, \qquad
  g_{\sM\sN} = \begin{pmatrix}
    0 & - n_\nu \\
    - n_\mu & h_{\mu\nu} + 2 n_{(\mu} B^{(v)}_{\nu)}
  \end{pmatrix},
\end{equation}
along with
\begin{equation}
  V_\sM = \begin{pmatrix} 0 \\ - n_\mu \end{pmatrix}, \quad
  v_{\sM} = \begin{pmatrix} -1 \\ B^{(v)}_\mu \end{pmatrix}, \quad
  g^{\sM\sN} = \begin{pmatrix}
    h^{\nu\r} B^{(v)}_\nu B^{(v)}_\r - 2 v^\mu B^{(v)}_\mu & h^{\nu\r}B^{(v)}_\r - v^\nu \\
    h^{\mu\nu} B^{(v)}_\nu - v^\mu & h^{\mu\nu}
  \end{pmatrix},
\end{equation}
such that
\begin{equation}
  n_\mu v^\mu = 1, \qquad 
  v^\mu h_{\mu\nu} = 0, \qquad
  n_\mu h^{\mu\nu} = 0, \qquad
  h_{\mu\r}h^{\r\nu} + n_\mu v^\nu = \d_{\mu}^{\ \nu}.
\end{equation}
The collection of fields $\{n_\mu,v^\mu,h^{\mu\nu},h_{\mu\nu},B^{(v)}_\mu\}$
defines a Newton-Cartan structure. The torsionless-ness condition
$H_{"sM"sN} = 0$ implies that the ``time-metric'' $n = n_\mu \df x^\mu$ is a
closed one-form, i.e.  $\df n = 0$; this is known to be true for torsionless
Newton-Cartan structures. Note that after choosing the said basis, the residual
diffeomorphisms are $x^\mu \ra x^\mu + \c^\mu(x^\nu)$ and
$x^- \ra \x^- + \c^- (x^\mu)$. The former of these are just the Newton-Cartan
diffeomorphisms, while the latter are known as ``mass gauge
transformations''. $B^{(v)}_\mu$ is the only field that transforms under this
gauge transformation
\begin{equation}
  \d_{\c^-} B^{(v)}_\mu = - \dow_\mu \c^-,
\end{equation}
and hence is known as the mass gauge field. The Levi-Civita connection
$\G^\sR_{\ \sM\sN}$ decomposes in this basis as
\begin{align}
  \label{NC_Connection}
  \G^{"l}_{\ \mu\nu}
  &= 
    v^{"l} \dow_{(\mu} n_{\nu)}
    + \half h^{"l\r} \lb \dow_\mu h_{\r\nu} + \dow_\nu h_{\r\mu} - \dow_\r h_{\mu\nu} \rb
    - "O^{(v)}_{\s(\mu} n_{\nu)} h^{"s"l}, \nn\\
  \G^{-}_{\ \mu\nu}
  &= 
    h_{"l(\mu} \tilde\N_{\nu)}v^{"l}
    - \tilde\N_{(\mu} B^{(v)}_{\nu)},
\end{align}
and all the remaining components zero. Here we have identified
$\G^{"l}_{\ "m"n}$ as the (torsionless) Newton-Cartan connection and denoted
the respective covariant derivative by $\tilde\N_\mu$. We have also defined the
(dual) frame vorticity as
\begin{equation}
  \O^{(v)}_{\mu\nu} = 2h_{\s[\nu}\tilde\N_{\mu]}v^\s = \dow_\mu B^{(v)}_\nu - \dow_\nu B^{(v)}_\mu.
\end{equation}
The covariant derivative $\tilde\N$ acts on the Newton-Cartan structure
appropriately
\begin{equation}
  \tilde\N_\mu n_\nu = 0, \qquad
  \tilde\N_\mu h^{\r\s} = 0, \qquad
  \tilde\N_\mu h_{\nu\r} = - 2 n_{(\nu} h_{\r)\s} \tilde\N_\mu v^\s.
\end{equation}
Note that $v^\sM$ was an arbitrary field chosen to perform the reduction, and
one is allowed to arbitrarily redefine it without changing the physics. This
leads to the invariance of the system under ``Milne transformations'' of the
Newton-Cartan structure
\begin{equation}\label{milne.trans}
  v^\mu \ra v^\mu + \p^\mu, \quad
  h_{\mu\nu} \ra h_{\mu\nu} - 2 n_{(\mu} \p_{\nu)} + n_\mu n_\nu \p^\r \p_\r, \quad
  B^{(v)}_\mu \ra B^{(v)}_\mu + \p_\mu - \half n_\mu \p^\r \p_\r,
\end{equation}
where $\p^\mu n_\mu = 0$, $\p_\mu = h_{\mu\nu}\p^\nu$. The fields $n_\mu$,
$h^{\mu\nu}$ and the connection $\G^\r_{\ \mu\nu}$ on the other hand are Milne
invariant.

We can now decompose the fluid velocity $u^\sM$ and the associated projector
$P^{\sM\sN}$ as
\begin{equation}
  u^\sM = \begin{pmatrix} u^\mu B_\mu \\ u^\mu \end{pmatrix}, \quad
  u_{\sM} = \begin{pmatrix} -1 \\ B_\mu \end{pmatrix}, \quad
  P_{\sM\sN} = \begin{pmatrix}
    0 & 0 \\
    0 & p_{\mu\nu}
  \end{pmatrix}, \quad
  P^{\sM\sN} = \begin{pmatrix}
    p^{\nu\r}B_\nu B_\r & p^{\mu\nu}B_\nu \\
    p^{\mu\nu} B_\nu & p^{\mu\nu}
  \end{pmatrix}.
\end{equation}
The fields $\{n_\mu, u^\mu, p^{\mu\nu},p_{\mu\nu},B_\mu\}$ define the Newton-Cartan structure
in the fluid frame of reference, satisfying,
\begin{equation}
  n_\mu u^\mu = 1, \qquad 
  u^\mu p_{\mu\nu} = 0, \qquad
  n_\mu p^{\mu\nu} = 0, \qquad
  p_{\mu\r}p^{\r\nu} + n_\mu u^\nu = \d_{\mu}^{\ \nu}.
\end{equation}
They can be re-expressed in terms of $\{n_\mu, v^\mu, h^{\mu\nu},h_{\mu\nu},B^{(v)}_\mu\}$ using
\cref{milne.trans} with $\p^\mu = \bar u^\mu = h^\mu{}_\nu u^\nu = u^\mu - v^\mu$,
\begin{equation}
  p^{\mu\nu} = h^{\mu\nu}, \quad
  p_{\mu\nu} = h_{\mu\nu} - 2 n_{(\mu} \bar u_{\nu)} + n_\mu n_\nu \bar u^\r \bar u_\r, \quad
  B_\mu = B^{(v)}_\mu + \bar u_\mu - \half n_\mu \bar u^\r \bar u_\r.
\end{equation}
The (dual) fluid vorticity is defined similar to the (dual) frame vorticity as,
\begin{equation}
  \O_{\mu\nu} = 2p_{\s[\nu}\tilde\N_{\mu]}u^\s = \dow_\mu B_\nu - \dow_\nu B_\mu.
\end{equation}

The mass current $\r^\mu$, energy current $\e^\mu$ and stress tensor
$t^{\mu\nu}$ on Newton-Cartan backgrounds can be respectively read out in terms
of $T^{\sM\sN}$ as (see~\cite{Banerjee:2015hra} for details)
\begin{equation}\label{current.redn}
  \r^\mu = -T^{\mu\sM}V_\sM, \qquad
  \e^\mu = -T^{\mu\sM}u_\sM, \qquad
  t^{\mu\nu} = P^{\mu}_{\ \sM} P^{\nu}_{\ \sN} T^{\sM\sN}.
\end{equation}
with $t^{\mu\nu} = t^{\nu\mu}$ and $t^{\mu\nu}n_\nu = 0$. They satisfy the
conservation laws
\begin{align}
  \text{Mass Conservation:} \qquad\qquad\qquad \ \tilde\N_\mu \r^{\mu}
  &= 0, \nn\\
  \text{Energy Conservation:}  \qquad\qquad\qquad \ \tilde\N_\mu \e^\mu
  &= 
    - \lb u^\mu \r^{\s} + t^{\mu\s} \rb p_{\s\nu} \tilde\N_\mu u^\nu, \nn\\
  \text{Momentum Conservation:} \ \ \tilde\N_\mu (u^\mu p^{\s}_{\ \nu} \r^{\nu} + t^{\mu\s})
  &= - \r^\mu \tilde\N_{\mu}u^\s.
\end{align}
Here, the energy current $\e^\mu$ and the stress tensor $t^{\mu\nu}$ in
\cref{current.redn} are defined in the fluid frame of reference; we can define
the respective quantities in an arbitrary frame of reference as
\begin{align}\label{changeOfGalileanFrame}
  \e^\mu_{(v)}
  &= -T^{\mu\sM}v_\sM
    = \e^\mu
    + u^\mu \bar u^\nu p_{\nu\r} \r^{\r}
    + \half \r^\mu \bar u^\r \bar u_\r
    + t^{\mu\nu} \bar u_\nu, \nn\\
  t^{\mu\nu}_{(v)}
  &= (P_{(v)})^{\mu}_{\ \sM} (P_{(v)})^{\nu}_{\ \sN} T^{\sM\sN}
    = t^{\mu\nu} + 2 \bar u^{(\mu}h^{\nu)}_{\ \ \s} \r^{\s} - \bar u^\mu \bar u^\nu
    \r^\s n_\s,
\end{align}
where $P_{(v)}^{\sM\sN} = g^{\sM\sN} + 2v^{(\sM}V^{\sN)}$. They satisfy the
respective conservation laws
\begin{align}
  \tilde\N_\mu \e_{(v)}^\mu
  &= 
    - \lb v^\mu \r^{\s} + t_{(v)}^{\mu\s} \rb h_{\s\nu} \tilde\N_\mu v^\nu \nn\\
  \tilde\N_\mu (v^\mu h^{\s}_{\ \nu} \r^{\nu} + t^{\mu\s}_{(v)})
  &= - \r^\mu \tilde\N_{\mu}v^\s.
\end{align}
In the following, we will only present the constitutive relations in the fluid
frame of reference. However, we can always use \cref{changeOfGalileanFrame} to
go to any arbitrary frame.

\subsection{Second order Galilean fluids}\label{2ndOrderGalilean}

Having reviewed the general rules of null reduction, we can now go on and reduce
the null fluid constitutive relations. To setup the notation, we have reduced
the first order fluid data in \cref{first-order-data-NC}. We can use the
formulae in \cref{current.redn} to convert the null fluid energy momentum tensor
$T^{"sM"sN}$ in \cref{tab:2nd-order-hydrostatic,tab:1der-C0,tab:0der-C1} into
Galilean fluid mass current $"r^{"m}$ in \cref{tab:NC-r}, energy current
$"e^{"m}$ in \cref{tab:NC-e} and stress tensor stress tensor $t^{"m"n}$ in
\cref{tab:NC-t}. The results schematically look like
\begin{equation}
  "r^{"m} = (R + "vs_{"r}) u^{"m} + "vs_{"r}^{"m}, \qquad
  "e^{"m} = (E + "vs_{"e}) u^{"m} + "vs_{"e}^{"m}, \qquad
  t^{"m"n} = P p^{"m"n} + "vs_t^{"m"n}.
\end{equation}
The tensors $"vs_{"r}$, $"vs_{"r}^{"m}$, $"vs_{"e}$, $"vs_{"e}^{"m}$ and
$"vs_t^{"m"n}$ contain derivative corrections with $"vs_t^{"m"n} = "vs_t^{"n"m}$
and $"vs_{"r}^{"m} n_{"m} = "vs_{"e}^{"m} n_{"m} = "vs_t^{"m"n} n_{"m} =
0$. These are the Galilean fluid constitutive relations in the offshell
hydrodynamic frame. If we are rather interested in the results in mass frame
defined in \cref{sec:massframe}, we can reduce \cref{offshellToMassFrame}
to get
\begin{equation}
  "vs_{"r,\text{mf}} = "vs_{"r,\text{mf}}^{"m} = "vs_{"e,\text{mf}} = 0, \qquad
  "vs_{"e,\text{mf}}^{"m} = "vs_{"e}^{"m} - \frac{E+P}{R} "vs_{"r}^{"m}, \qquad
  "vs_{t,\text{mf}}^{"m"n} = "vs_t^{"m"n}
  - p^{"m"n}\lb\frac{\dow P}{\dow E} "vs_{"e} + \frac{\dow P}{\dow R} "vs_{"r}\rb.
\end{equation}
These mass frame results have also been presented in
\cref{tab:NC-r,tab:NC-e,tab:NC-t} alongside their offshell frame counterparts.

\begin{table}[!p]
    \renewcommand{\arraystretch}{1.6}
  \centering
  \begin{tabular}[t]{cccccc}
    \midrule\midrule
    \multicolumn{2}{c}{Null backgrounds}
    & \multicolumn{2}{c}{Newton-Cartan}
    & \multicolumn{2}{c}{Non-covariant (flat)} \\
    \midrule\midrule
    
    \multicolumn{6}{c}{Non-hydrostatic --- onshell independent} \\
    \midrule
    $"Q$
    & $\N_{"sM} u^{"sM}$
    & $"Q$
    & $\tilde\N_{"m} u^{"m}$
    & $"Q$
    & $\dow_{i} u^{i}$ \\
    
    $"t^{"sM}$, $\bar{"t}^{"sM}$
    & $P^{"sM"sN}\frac{1}{T} \dow_{"sN} T$
    & $"t^{"m}$, $\bar{"t}^{"m}$
    & $p^{"m"n}\frac{1}{T} \dow_{"n} T$
    & $"t^{i}$, $\bar{"t}^{i}$
    & $\frac{1}{T} \dow^i T$ \\
    
    $"s^{"sM"sN}$
    & $2 P^{"sR\langle"sM}P^{"sN\rangle}{}_{"sS} \N_{"sR} u^{"sS}$
    & $"s^{"m"n}$
    & $2 p^{"r\langle"m} \tilde\N_{"r} u^{"n\rangle}$
    & $"s^{ij}$
    & $2 \dow^{\langle i} u^{j\rangle}$ \\
    
    \midrule\midrule
    \multicolumn{6}{c}{Non-hydrostatic --- onshell dependent} \\
    \midrule

    $"Q_{T}$
    & $u^{"sM} \frac1T \dow_{"sM} T$
    & $"Q_{T}$
    & $u^{"m} \frac1T \dow_{"m} T$
    & $"Q_{T}$
    & $\frac1T \lb \dow_t T + u^i \dow_i T \rb$ \\
    
    $"Q_{"vp}$
    & $u^{"sM} T \dow_{"sM} "vp$
    & $"Q_{"vp}$
    & $u^{"m} T \dow_{"m} "vp$
    & $"Q_{"vp}$
    & $T \lb \dow_t "vp + u^i \dow_i "vp \rb$ \\

    $"t^{"sM}_{"vp}$
    & $P^{"sM"sN}\lb T \dow_{"sN} "vp + u^{"sR}\N_{"sR}u_{"sN} \rb$
    & $"t^{"m}_{"vp}$
    & $ p^{"m"n} T \dow_{"n} "vp + u^{"sR}\N_{"sR}u^{"m}$
    & $"t^{i}_{"vp}$
    & $T \dow^i "vp + \dow_t u^{i} + u^j\dow_j u^i$ \\

    \midrule\midrule
    \multicolumn{6}{c}{Hydrostatic} \\
    
    \midrule

    $\bar{"t}^{"sM}_{"vp}$
    & $P^{"sM"sN}T\dow_{"sN}"vp$
    & $\bar{"t}^{"m}_{"vp}$
    & $p^{"m"n}T\dow_{"n}"vp$
    & $\bar{"t}^{i}_{"vp}$
    & $T\dow^i "vp$ \\

    $"o^{"sM"sN}$
    & $2 P^{"sR["sM}P^{"sN]"sS} \N_{"sR} u_{"sS}$
    & $"o^{"m"n}$
    & $2 p^{"r["m} \tilde\N_{"r} u^{"n]}$
    & $"o^{ij}$
    & $2 "dow^{[i}u^{j]}$ \\

    \midrule\midrule
  \end{tabular}
  \caption{\label{first-order-data-NC} First order fluid data in null
    background, Newton-Cartan and non-covariant notations. Non-covariant results
    have been specialised to flat backgrounds.}
\end{table}

\begin{table}[!p]
  \renewcommand{\arraystretch}{2}
  \centering
  \begin{tabular}[h]{ccc}
    \midrule\midrule
    \#
    & \parbox[c]{0.4\textwidth}{\centering $"r^{"m}$}
    & \parbox[c]{0.4\textwidth}{\centering $"r_{\text{mf}}^{\mu}$} \\
    \midrule\midrule

    $P$
    & \multicolumn{2}{c}{$\dsp \frac{1}{T}\frac{\dow P}{\dow "vp}~u^{"m} \equiv R u^{"m}$} \\
    
    \midrule

    $"a_1$
    & $\dsp \frac{1}{T} \lb \half \frac{\dow"a_1}{\dow"vp}
      \bar{"t}_{"vp}^{"m} \bar{"t}_{"vp}{}_{"m}
      - \tilde\N_{"n}("a_1 T \bar{"t}_{"vp}^{"n}) \rb u^{"m}$
    & $0$ \\

    $"a_2$
    & $\dsp \frac{1}{T} \lb \frac{\dow"a_2}{\dow"vp}
      \bar{"t}_{"vp}^{"m} \bar{"t}_{"m}
      - \tilde\N_{"n}("a_2T \bar{"t}^{"n}) \rb u^{"m} $
    & $0$ \\

    $"a_3$
    & $\dsp
      \frac1{4T} \frac{\dow"a_3}{\dow"vp} "o^{"r"s}"o_{"r"s}~u^{"m}
      - \tilde\N_{"n}\lb "a_3 "o^{"n"m} \rb$
    & $0$ \\

    $"a_5$
    & $\dsp \frac1{T} \frac{\dow"a_5}{\dow"vp} R u^{"m}$
    & $0$ \\
    
    \midrule\midrule

  \end{tabular}
  \caption{\label{tab:NC-r} Mass current of a Galilean fluid up to second
    derivative order in offshell frame and mass frame respectively. Note that in
    mass frame, the mass current is simply $"r^{"m}_{\text{mf}} = R u^{"m}$.}
\end{table}

\begin{table}[!p]
  \renewcommand{\arraystretch}{2}
  \centering
  \begin{tabular}[h]{ccc}
    \midrule\midrule
    \#
    & \parbox[c]{0.44\textwidth}{\centering $"e^{"m}$}
    & \parbox[c]{0.44\textwidth}{\centering $"e^{"m}_{\text{mf}}$} \\
    \midrule\midrule

    $P$
    & \multicolumn{2}{c}{$\dsp T^2\frac{\dow (P/T)}{\dow T} u^{"m} \equiv E u^{"m}$} \\
    \midrule

    $"k T$
    & \multicolumn{2}{c}{$- T"k"t^{"m}$} \\

    \midrule
    
    $"a_1$
    & $\dsp \half \frac{\dow(T "a_1)}{\dow T}
      \bar{"t}_{"vp}^{"r} \bar{"t}_{"vp}{}_{"r}~u^{"m}
      + "a_1 \bar{"t}^{"m}_{"vp} "Q_{"vp}$
    & $\dsp 
      "a_1 \bar{"t}^{"m}_{"vp} "Q_{"vp}$ \\[2ex]

    $"a_2$
    & \parbox[c]{0.44\textwidth}{\centering
      $\dsp \lB T^2 \frac{\dow ("a_2/T)}{\dow T} \bar{"t}_{"vp}^{"r} \bar{"t}_{"r}
      - T \N_{"n}\lb\frac{"a_2}{T} \bar{"t}^{"n}_{"vp} \rb \rB u^{"m}$
      $+ "a_2 \bar{"t}^{"m}  "Q_{"vp}
      + "a_2 \bar{"t}^{"m}_{"vp} "Q_{T}$}
    & $"a_2 \bar{"t}^{"m}  "Q_{"vp}
      + "a_2 \bar{"t}^{"m}_{"vp} "Q_{T}$ \\[3ex]

    $"a_3$
    & $\dsp \frac14 T^2 \frac{\dow ("a_3/T)}{\dow T} "o^{"n"r}"o_{"n"r}~u^{"m}
      + "a_3 "o^{"m"n}a_{"n}$
    & $\dsp "a_3 "o^{"m"n}a_{"n}
      + \frac{E+P}{R} \tilde\N_{"n}\lb "a_3 "o^{"n"m} \rb $ \\

    $"a_4$
    & $- \tilde\N_{"n}("a_4"o^{"n"m})$
    & $- \tilde\N_{"n}("a_4"o^{"n"m})$ \\[2ex]

    $"a_5$
    & \parbox[c]{0.44\textwidth}{\centering
      $\dsp \lb T^2 \frac{\dow ("a_5/T)}{\dow T} R
      + 2 p^{"r"s}\tilde\N_{"r}\tilde\N_{"s}"a_5 \rb u^{"m}$\\
      $+ 2 "a_5 p^{"m"n}R_{"n"r} u^{"r}
    - 2 p^{"m"n} u^{"r} \tilde\N_{"n} \tilde\N_{"r}"a_5 $}
    & $2 "a_5 p^{"m"n}R_{"n"r} u^{"r}
      - 2 p^{"m"n} u^{"r} \tilde\N_{"n} \tilde\N_{"r}"a_5 $ \\
    
    \midrule
      
    $"d^+_1$ & \multicolumn{2}{c}{$-\half "d^+_1 "s^{\mu\nu}"t_{"n}$} \\
    $"d^+_2$ & \multicolumn{2}{c}{$- "d^+_2 "t^{"m}"Q$} \\
    \midrule
    
    $"d^+_6$ & \multicolumn{2}{c}{$- \half "d^+_6 "s^{"m"n} \bar{"t}_{"vp}{}_{"n}$} \\
    $"d^+_7$ & \multicolumn{2}{c}{$- "d^+_7 \bar{"t}^{"m}_{"vp} "Q$} \\
    $\bar{"d}_8$ & \multicolumn{2}{c}{$-\half\bar{"d}_8 "o^{"m"r}"t_{"r}$} \\
    \midrule
    
    $"d^+_{10}$
    & \multicolumn{2}{c}{$- p^{"m"r} \lB
      "d^+_{10}\dow_{"r}"Q + \frac{T}2 "Q \dow_{"r}\bfrac{"d^+_{10}}{T}\rB
      + \half "d^+_{10} p^{"m"n} "t_{"s} \tilde\N_{"n} u^{"s} $} \\
    $"d^-_{10}$
    & \multicolumn{2}{c}{$- \half "d^-_{10} "t^{"s} \tilde\N_{"s} u^{"m}$} \\

    $"d^+_{11}$
    & \multicolumn{2}{c}{$- \half \lB
        "d^+_{11}\tilde \N_{"n} "s^{"m"n}
        + \frac{T}2 "s^{"m"n} \dow_{"n}\bfrac{"d^+_{11}}{T} \rB
      + \frac14 "d^+_{11} \lb
        "t^{"n} \tilde \N_{"n} u^{"m}
        + "t^{"m} "Q
        - \frac{2}{d} p^{"m"r} "t_{"s} \tilde\N_{"r} u^{"s}
      \rb $} \\

    $"d^-_{11}$
    & \multicolumn{2}{c}{$- \frac14 "d^-_{11} \lb
      p^{"m"r} "t_{"s} \tilde\N_{"r} u^{"s}
        + "t^{"m} "Q 
      - \frac{2}{d} "t^{"n}\tilde \N_{"n} u^{"m} \rb$} \\

    $\bar{"d}_{12}$
    & \multicolumn{2}{c}{$- \bar{"d}_{12} a^{"m} "Q$} \\

    $\bar{"d}_{13}$
    & \multicolumn{2}{c}{$- \half \bar{"d}_{13} "s^{"m"n}a_{"n}$} \\
    
    \midrule\midrule

  \end{tabular}
  \caption{\label{tab:NC-e} Energy current of a Galilean fluid up to second
    derivative order in offshell frame and mass frame respectively. Note that
    the only difference in the two frames comes in the hydrostatic sector.}
\end{table}

\begin{table}[!p]
  \renewcommand{\arraystretch}{1.5}
  \centering
  \begin{tabular}[h]{ccc}
    \midrule
    \#
    & \parbox[c]{0.3\textwidth}{\centering $t^{"m"n}$}
    & \parbox[c]{0.58\textwidth}{\centering $t^{"m"n}_{\text{mf}}$} \\
    \midrule\midrule

    $P$
    & \multicolumn{2}{c}{$\dsp P p^{"m"n}$} \\
    \midrule

    $"h$
    & \multicolumn{2}{c}{$-"h "s^{"m"n}$} \\

    $"z$
    & \multicolumn{2}{c}{$-"z p^{"m"n} "Q$} \\

    \midrule

    $"a_1$
    & $\dsp \half "a_1 p^{"m"n} \bar{"t}_{"vp}^{"r} \bar{"t}_{"vp}{}_{"r}
      - "a_1 \bar{"t}_{"vp}^{"m} \bar{"t}_{"vp}^{"n}$
    & \parbox[c]{0.58\textwidth}{\centering
      $\dsp \half "a_1 p^{"m"n} \bar{"t}_{"vp}^{"r} \bar{"t}_{"vp}{}_{"r}
      - "a_1 \bar{"t}_{"vp}^{"m} \bar{"t}_{"vp}^{"n}
      - p^{"m"n}\frac{\dow P}{\dow E} \half \frac{\dow(T "a_1)}{\dow T}
      \bar{"t}_{"vp}^{"r} \bar{"t}_{"vp}{}_{"r} $
      $\dsp - p^{"m"n} \frac{\dow P}{\dow R} \frac{1}{T} \lb
      \half \frac{\dow"a_1}{\dow"vp} \bar{"t}_{"vp}^{"m} \bar{"t}_{"vp}{}_{"m}
      - \tilde\N_{"n}("a_1 T \bar{"t}_{"vp}^{"n}) \rb $}\\[4ex]

    $"a_2$
    & $p^{"m"n}"a_2 \bar{"t}_{"vp}^{"r} \bar{"t}_{"r}
      - 2"a_2 \bar{"t}_{"vp}^{("m} \bar{"t}^{"n)}$
    & \parbox[c]{0.58\textwidth}{\centering
      $\dsp p^{"m"n}"a_2 \bar{"t}_{"vp}^{"r} \bar{"t}_{"r}
      - 2"a_2 \bar{"t}_{"vp}^{("m} \bar{"t}^{"n)}$
      $\dsp - p^{"m"n} \frac{\dow P}{\dow E} \lb
      T^2 \frac{\dow ("a_2/T)}{\dow T} \bar{"t}_{"vp}^{"r} \bar{"t}_{"r}
      - T \N_{"n}\lb\frac{"a_2}{T} \bar{"t}^{"n}_{"vp} \rb \rb$
      $\dsp - p^{"m"n} \frac{\dow P}{\dow R} \lb \frac{\dow"a_2}{\dow"vp}
      \bar{"t}_{"vp}^{"r} \bar{"t}_{"r}
      - \tilde\N_{"n}("a_2T \bar{"t}^{"n}) \rb$} \\[6ex]

    $"a_3$
    & $\dsp \frac14 "a_3 p^{"m"n} "o^{"r"s}"o_{"r"s}
      - "a_3 "o^{"m}{}_{"r} "o^{"n"r}$
    & \parbox[c]{0.58\textwidth}{\centering
      $\dsp \frac14 "a_3 p^{"m"n} "o^{"r"s}"o_{"r"s}
      - "a_3 "o^{"m}{}_{"r} "o^{"n"r}$
      $\dsp
      - p^{"m"n} \frac{\dow P}{\dow E} \frac{T^2}{4}
      \frac{\dow ("a_3/T)}{\dow T} "o^{"n"r}"o_{"n"r}
      - p^{"m"n} \frac{\dow P}{\dow R} \frac1{4T}
      \frac{\dow"a_3}{\dow"vp} "o^{"r"s}"o_{"r"s}$} \\[5ex]

    $"a_5$
    & \parbox[c]{0.3\textwidth}{\centering
      $p^{"m"n} \lb "a_5 R  - 2 p^{"r"s}\tilde\N_{"r}\tilde\N_{"s}"a_5 \rb$
      $- 2 p^{"m"r} p^{"n"s} \lb "a_5 R_{"r"s} - \tilde\N_{"r}\tilde\N_{"s}"a_5\rb$}
    & \parbox[c]{0.59\textwidth}{\centering
      $p^{"m"n} \lb "a_5 R  - 2 p^{"r"s}\tilde\N_{"r}\tilde\N_{"s}"a_5 \rb
      - 2 p^{"m"r} p^{"n"s} \lb "a_5 R_{"r"s} - \tilde\N_{"r}\tilde\N_{"s}"a_5\rb$
      $\dsp - p^{"m"n} \frac{\dow P}{\dow E} \lb
      T^2 \frac{\dow ("a_5/T)}{\dow T} R
      + 2 p^{"r"s}\tilde\N_{"r}\tilde\N_{"s}"a_5 \rb$
      $\dsp - p^{"m"n} \frac{\dow P}{\dow R}
      \frac1{T} \frac{\dow"a_5}{\dow"vp} R$} \\

    \midrule
   
    $"d^-_1$
    & \multicolumn{2}{c}{$-"d^+_1 "t^{\langle"m}"t^{"n\rangle}$} \\
    $"d^-_2$
    & \multicolumn{2}{c}{$- "d^-_2 p^{"m"n} "t^{"r}"t_{"r}$} \\
    $"d^+_3$
    & \multicolumn{2}{c}{$- 2"d^+_3 "s^{"m"n} "Q$} \\
    $"d^-_3$
    & \multicolumn{2}{c}{$- "d^-_3 p^{"m"n} "s^{"r"s}"s_{"r"s}$} \\
    $"d_4$
    & \multicolumn{2}{c}{$-"d_4 "s^{\langle"m"r}"s_{"r}{}^{"n\rangle}$} \\
    $"d_5$
    & \multicolumn{2}{c}{$-2"d_5 p^{"m"n}"Q^2$} \\
    \midrule

    $"d^-_6$
    & \multicolumn{2}{c}{$-"d^+_6 \bar{"t}^{\langle"m}_{"vp} "t^{"m\rangle}$} \\
    
    $"d^-_7$
    & \multicolumn{2}{c}{$- "d^-_7 p^{"m"n} \bar{"t}^{"r}_{"vp} "t_{"r}$} \\
    
    $\bar{"d}_9$
    & \multicolumn{2}{c}{$-\bar{"d}_9 "o^{\langle"m"r}"s_{"r}{}^{"n\rangle}$} \\
    \midrule

    $"d^-_{10}$
    & \multicolumn{2}{c}{$- p^{"m"n} \lB "d^-_{10} \tilde\N_{"r}"t^{"r}
      + \frac{T}2 "t^{"r}\dow_{"r} \bfrac{"d^-_{10}}{T} \rB$} \\

    $"d^-_{11}$
    & \multicolumn{2}{c}{$- P^{"r\langle"m}P^{"n\rangle}{}_{"s} \lB
        "d^-_{11} \tilde\N_{"r}"t^{"s}
      + \frac{T}2 "t^{"sR}\dow_{"sT}\bfrac{"d^-_{11}}{T} \rB$} \\

    $\bar{"d}_{12}$
    & \multicolumn{2}{c}{$- 2p^{"m"n} \lB \bar{"d}_{12} u^{"r}\dow_{"r}"Q + \frac{T}{2}
      "Q u^{"r} \dow_{"r}\bfrac{\bar{"d}_{12}}{T} \rB
      - p^{"m"n} \bar{"d}_{12}\lb "Q^2 - a^{"r}"t_{"r}\rb$} \\

    $\bar{"d}_{13}$
    & \multicolumn{2}{c}{$\dsp - \lB
      \bar{"d}_{13} u^{"r}\tilde\N_{"r} "s^{"m"n}
      + \frac{T}{2} "s^{"m"n} u^{"r}\dow_{"r}\bfrac{\bar{"d}_{13}}{T} \rB$
      $\dsp
      - \bar{"d}_{13}
      \lb\frac{1}{2} "s^{"m"n}"Q - a^{\langle"m}"t^{"n\rangle}\rb$} \\

    \midrule
    
  \end{tabular}
  \caption{\label{tab:NC-t} Stress-energy tensor of a Galilean fluid up to
    second derivative order in offshell frame and mass frame respectively. Note
    that the only difference in the two frames comes in the hydrostatic
    sector.}
\end{table}

\paragraph*{Summary of transport coefficients:} At ideal order, there is just
one independent transport coefficient: the thermodynamic pressure
$P$. Thermodynamic energy density $E$ and mass density $R$ are determined in
terms of $P$ via the thermodynamic relation
\begin{equation}
  "dd P = \frac{E+P}{T}"dd T + TR "dd "vp.
\end{equation}
At first order, there are three new transport coefficients: shear viscosity
$"h$, bulk viscosity $"z$ and thermal conductivity $"k$. All three of them are
dissipative and are required to be non-negative by the second law. At second
order, there are 25 independent transport coefficients. 5 of them are hydrostatic belonging
to Class H$_{\text{S}}$
\begin{equation}
  "a_1, \qquad "a_2, \qquad "a_3, \qquad "a_4, \qquad "a_5.
\end{equation}
Other 20 are non-hydrostatic. 11 in Class $\overline{\text{D}}$
\begin{equation}
  \bar{"d}_1, \qquad \bar{"d}_2, \qquad \bar{"d}_3, \qquad \bar{"d}_6, \qquad
  \bar{"d}_7, \qquad \bar{"d}_8, \qquad \bar{"d}_9, \qquad \bar{"d}_{10}, \qquad
  \bar{"d}_{11}, \qquad \bar{"d}_{12}, \qquad \bar{"d}_{13},
\end{equation}
and 9 in Class D
\begin{equation}
  "d_1,\qquad "d_2,\qquad "d_3,\qquad "d_4,\qquad "d_5,\qquad "d_6,\qquad
  "d_7,\qquad "d_{10},\qquad "d_{11}.
\end{equation}
Second law gives no inequality constraints on these transport coefficients. For
some transport coefficients in \cref{tab:NC-r,tab:NC-e,tab:NC-t}, we have
denoted $"d^{\pm}_i = ("d \pm \bar{"d}_i)/2$.

\subsection{Non-covariant results in flat spacetime}

When the Galilean fluid is coupled to a flat background, it is perhaps more
fitting to express the results in the conventional non-covariant notation where
the time and space indices are treated distinctly. To make the transition, we
note that on a Newton-Cartan background, we can choose a basis
$\{x^\mu\} = \{t,x^i\}$ such that the Galilean frame velocity
$(v^\mu) = \dow_t$. In this basis, we decompose the Newton-Cartan structure as
\begin{equation}
  n_\mu = \begin{pmatrix} 1 \\ 0 \end{pmatrix}, \quad
  v^\mu = \begin{pmatrix} 1 \\ 0 \end{pmatrix}, \quad
  p^{\mu\nu} = \begin{pmatrix}
    0 & 0 \\ 0 & \d^{ij}
  \end{pmatrix}, \quad
  p_{\mu\nu} = \begin{pmatrix}
    0 & 0 \\ 0 & \d_{ij}
  \end{pmatrix}, \quad
  B^{(v)}_\mu = 0,
\end{equation}
where $\d^{ij} = \d_{ij}$ is the Kronecker delta. It can be checked that the
respective Newton-Cartan connection $\G^{"l}_{\ "m"n} = 0$, justifying the
spacetime to be flat. The Newton-Cartan structure in the fluid frame can also be
worked out from here to be
\begin{equation}
  u^\mu = \begin{pmatrix} 1 \\ u^i \end{pmatrix}, \quad
  B_\mu = \begin{pmatrix} -\half u^k u_k \\ u_i \end{pmatrix}, \quad
  p^{\mu\nu} = \begin{pmatrix}
    0 & 0 \\ 0 & \d^{ij}
  \end{pmatrix}, \quad
  p_{\mu\nu} = \begin{pmatrix}
    u^k u_k & - u_j \\ - u_i & \d_{ij}
  \end{pmatrix}.
\end{equation}
We have enlisted the one derivative fluid data in \cref{first-order-data-NC} to
aid the transition of the constitutive relations to non-covariant notation.

In flat spacetime, the conservation laws take the well known form
\begin{align}\label{noncov-conservation}
  \text{Mass Conservation:} \qquad \dow_t \r^t + \dow_i \r^i
  &= 0 \nn\\
  \text{Energy Conservation:} \ \ \dow_t \e_{(v)}^t + \dow_i \e_{(v)}^i
  &= 0 \nn\\
  \text{Momentum Conservation:} \quad \ \dow_t \r^{j}
  + \dow_i t^{ij}_{(v)}
  &= 0.
\end{align}
Here we have identified various Galilean quantities as expressed in the Galilean
frame defined by $\dow_t$: mass density $\r^t$, mass current $\r^i$, energy
density $\e^t_{(v)}$, energy current $\e^i_{(v)}$ and stress tensor
$t^{ij}_{(v)}$. For a Galilean fluid, they take a schematic form
\begin{align}
  "r^t
  &= R + "vs_{"r}, \qquad
    "r^i
    = "r^t u^i + "vs_{"r}^i, \qquad
    t^{ij}_{(v)}
    = P "d^{ij} + "r^t u^i u^j
    + 2 u^{(i}"vs_{"r}^{j)} + "vs_t^{ij}, \nn\\
  \e^t_{(v)}
  &= E 
    + \half (R + "vs_{"r}) u^k u_k
    + "vs_{"r}^k u_k
    + "vs_{"e}, \qquad
  \e^i_{(v)}
  = \lb \e^t_{(v)} + P \rb u^i
    + \half "vs_{"r}^i u^k \bar u_k
    + "vs_{"e}^i
    + "vs_t^{ij}u_j,
\end{align}
where $"vs_{"r}$, $"vs^i_{"r}$, $"vs_{"e}$, $"vs^i_{"e}$ and $"vs^{ij}_{t}$
contain all the derivative corrections. We can also use the mass hydrodynamic frame
\begin{align}
  "r^t_{\text{mf}}
  &= R, \qquad
    "r^i_{\text{mf}} = R u^i, \qquad
    t^{ij}_{(v),\text{mf}} = P "d^{ij} + R u^i u^j  + "vs_{t,\text{mf}}^{ij}, \nn\\
  \e^t_{(v),\text{mf}}
  &= E 
    + \half R u^k u_k, \qquad
  \e^i_{(v),\text{mf}}
  = \lb E + P + \half R u^k u_k \rb u^i
    + "vs_{"e,\text{mf}}^i
    + "vs_{t,\text{mf}}^{ij}u_j.
\end{align}
in which the expressions look most familiar. For the constitutive relations up
to second order, the respective derivative corrections can be directly read out
from \cref{tab:NC-r,tab:NC-e,tab:NC-t} using results in \cref{first-order-data-NC}.

We have now completed the most generic analysis of the constitutive relations of
a Galilean fluid up to second derivative order. Before closing this paper, in
the next section we present an example of how these second order terms might
find relevance in a physical process. We consider a ball being dragged through a
Galilean fluid and study corrections to the Stokes' law due to a representative
second order term.

\section{Second order corrections to the Stokes' law}\label{sec:application}

In this section, we analyse the effect of the second order transport
coefficients on a well known hydrodynamic phenomenon, namely the Stokes'
law.\footnote{This is a very famous problem and described in many texts on fluid
  dynamics.}  The physical setup is the following: we are interested in finding
out the fluid profile around a spherical ball that is moving at a constant
velocity in a fluid of infinite extent. Given the symmetries of the problem, we
will choose a spherical coordinate system $(r,"q,"f)$.  The radius of the ball
$\ell$, its constant velocity $U\hat z$ with respect to the fluid at infinity,
fluid density $R$ and fluid viscosity $"h$ are the parameters of the
problem. The fluid resists the motion of the ball due to its viscosity by
applying a drag force opposite to its direction of motion. Its magnitude is
determined by the Stokes' law
\begin{equation}\label{SL}
  \vec F = - 6\pi\eta \ell U \hat z.
\end{equation}
This law follows from the momentum conservation \cref{noncov-conservation},
commonly known as the Navier-Stokes (NS) equation, under certain assumptions
which we outline below. For a first order dissipative non-relativistic fluid NS
equation takes the form
\begin{equation}\label{NS}
  R\partial_t u^i  + R u^{j}\partial_j u^{i} =
  - \partial^i P   +
  \eta \partial^2 u^i  + \lb\zeta+\frac{1}{3}\eta\rb\partial^i \partial_j u^j, 
\end{equation}
where the transport coefficients $"h$, $"z$ have been assumed to be constant in
spacetime. The problem can be made time independent by working in the rest
frame of the ball, so that far away from the ball, the velocity of fluid is
$-U\hat{z}$. We will ignore any inertial effects all together. Next we consider
the fluid to be incompressible ($R$ is constant) and hence by virtue of the
equation of continuity (mass conservation equation), the fluid velocity becomes
divergenceless
\begin{equation}
  \dow_t R + \dow_i (R u^i) = R \dow_i u^i = 0.
\end{equation}
These assumptions simplify the NS equation \cref{NS} to be
\begin{equation}\label{NS-assum}
  R u^{j}\partial_j u^{i} = - \partial^i P   + \eta \partial^2 u^i.
\end{equation}
Due to the axisymmetric nature of our problem, $u^{"f}$ can be taken to be $0$
and all the other fields to be independent of the $"f$ coordinate. With this
ansatz, the fluid velocity can be expressed in terms of a \emph{stream function}
$"y(r,"q)$
\begin{equation}\label{vel-ansatz}
  \vec u = \frac{1}{r^2\sin\theta}
  \lb \partial_\theta \psi ~\hat r - \partial_r \psi ~\hat{"q} \rb.
\end{equation}
So, for a given fluid (characterised by $\rho, \eta$), the problem is reduced to
solving \cref{NS-assum} for the pressure $P$ and the stream function $"y$, with
boundary conditions: (1) P = $P_0$ is constant at infinity, (2)
$\vec u = -U\hat{z}$ at infinity and (3) $\vec u$ = 0 at the surface of the
ball. The solutions are given as
\begin{equation}\label{stokes-solutions}
  "y = \half U \sin^2\q \lb \frac{\ell^3}{2r} - \frac{3}{2}\ell r + r^2 \rb, \qquad
  P = P_0 - "h \ell \frac{3U\cos"q}{2r^2}.
\end{equation}
We define the force per unit area on the ball as
\begin{equation}\label{forceDifferential}
  \df \vec F = t^{ij} \hat x_i \df a_j
  = \ell^2 \sin"q ~ \df"f \df "q \lb P ~ \hat r - "h "s^{r"q} \hat{"q} \rb 
  = \ell^2 \sin"q ~ \df"f \df "q \lb P_0 ~ \hat r - \frac{3"h U}{2\ell} \hat z \rb.
\end{equation}
Stokes' law follows from here by integrating this equation over the surface of
the ball.

\subsection{Navier-Stokes equation with second order scalar corrections}

We now study the Navier-Stoke equation \bref{NS} in presence of second order
transport coefficients. In particular, we want to see the second order
corrections to the Stokes' law\footnote{We thank Suvankar Dutta for suggesting
  this application.} given in \cref{SL}.  The full analysis of \cref{NS} in
presence of all second order coefficients is naturally very involved, and is out
of the scope of this paper. We plan to return to this analysis in the
future. Here, we are only interested in the second order scalar terms. As it
turns out, such terms do not change the fluid velocity profile and only affect
the pressure.

We modify the stress tensor of our non-relativistic fluid with a scalar term
proportional to a second order non-hydrostatic data $S$, giving us
\begin{equation}\label{t-withS}
  t^{ij} = R u^i u^j + P "d^{ij} - \eta\sigma^{ij} + \tilde{\eta} S "d^{ij},
\end{equation}
leaving the energy and mass currents unmodified. $\tilde{"h}$ is the associated
transport coefficient. In terms of the transport
coefficients defined in \cref{tab:NC-t}, $\tilde{"h}$ can be a linear
combination of\footnote{The combinations of transport coefficients are
  specifically chosen, so that the energy current does not get any second order
  corrections.}
\begin{equation}
  "d_2^-, \qquad
  "d_3^-, \qquad
  "d_5, \qquad
  "d_7^-, \qquad
  "d_{10}^- - \half "d_1^+ - \frac1{2d}"d_2^+ + \half \bar{"d}_8, \qquad
  \bar{"d}_{12} + \frac{E+P}{R}"d_2^+ + "d_7^+.
\end{equation}
Under the assumptions of incompressibility and constant transport coefficients,
the associated NS equation takes the form
\begin{equation}\label{NScor}
  \partial^i P + R u^{j}  \partial_j u^{i}  - \eta \partial^2 u^i  +
  \tilde{\eta}\partial^i S  = 0. 
\end{equation}
Taking a curl of this equation, $P$, $S$ and $\tilde{"h}$ drop out of the
equation. The resultant equation is just an equation in the velocity ignorant of
our $S$ corrections. Together with the divergence-less condition, it completely
determines the velocity profile which is independent of $S$ and is simply given
by \cref{vel-ansatz}.

Once we have obtained the velocity profile, the pressure $P$ can be obtained by
solving \cref{NScor}. In general, the solutions will crucially depend on the
form of $S$. For example, if $S$ only involves the velocity and its derivatives,
\cref{NScor} simply becomes a homogeneous first order differential equation for
$P$. The solution is given by a trivial extension of \cref{stokes-solutions}
\begin{equation}
  P = P_0 - "h \ell \frac{3U\cos"q}{2r^2} - \tilde{"h} S,
\end{equation}
where $S$ is evaluated on the velocity profile. Since the stress tensor in
\cref{t-withS} only depends on the combination $P+\tilde{"h} S$, we can see that
the contributions from $\tilde{"h}$ drop out of it after plugging in the
solutions. It follows from \cref{forceDifferential} therefore, that the Stokes'
law does not receive any corrections. In fact the same argument goes through for
any arbitrary $S$. Note that \cref{NScor} can be rewritten as
\begin{equation}
  \partial^i \lb P + \tilde{"h} S\rb
  = - R u^{j}  \partial_j u^{i} + \eta \partial^2 u^i.
\end{equation}
After plugging in the solution for velocity, this equation can be integrated
once to give
\begin{equation}
  P + \tilde{"h} S = P_0 - "h \ell \frac{3U\cos"q}{2r^2}.
\end{equation}
Depending on the pressure dependence of $S$, this equation might be non-trivial
to solve. However, as far as $t^{ij}$ is concerned, we are only interested in
the combination $P + \tilde{"h}S$. It follows therefore, that $t^{ij}$ and hence
the Stokes' law does not receive any corrections due to $\tilde{"h}$.

We have argued that the non-hydrostatic second order scalar corrections $S$ do
not affect the Stokes' law under the assumptions of incompressibility and
constant transport coefficients (which are the same as imposed by Landau in his
book~\cite{landau1959fluid}). To see a non-trivial effect on the Stokes' law
therefore, we should consider other tensorial corrections to the stress
tensor. The full analysis however, is expected to be pretty involved
analytically, and we intend to return to this in the near future.

\section{Discussion}\label{sec:discussion}

In this paper, we have performed a complete second order analysis of uncharged
parity-even non-relativistic hydrodynamics using null fluid formalism. Second
order terms in relativistic hydrodynamics are required to maintain
causality. Although there is no notion of causality in a non-relativistic
theory, if we look at it as a non-relativistic limit of a relativistic theory,
we might expect to see some signatures of the causality requirement. It is
therefore important to study the effect of second order terms in
non-relativistic fluid constitutive relations. The respective results can be
found in \cref{2ndOrderGalilean} in \cref{tab:NC-r,tab:NC-e,tab:NC-t}. To
summarise, there are 25 new transport coefficients that appear at second
order. 5 of them are hydrostatic, i.e.\ they determine the equilibrium
configuration of the fluid. 9 others are dissipative, i.e.\ they are responsible
for the production of entropy during dynamical processes, while the remaining 11
quantify dynamical processes which do not cause dissipation.

To understand the physical effect of these second order terms, we also explore
how some of these might modify the well known Stokes' law, which tells us the
drag force experienced by a body while moving through a fluid. To first order,
we already know that it is proportional to the shear viscosity of the fluid
\cite{landau1959fluid}. We concluded that non-hydrostatic scalar corrections to
the stress tensor (terms that appear in the stress tensor as
$t^{ij}\sim \tilde{"h}S "d^{ij}$) do not affect the Stokes' law at all. This
accounts for 6 out of 25 transport coefficients. There are 7 other terms which
only affect the energy profile and hence cannot contribute to the drag
force. The remaining 12 coefficients can in principle however, affect the
Stokes' law in a non-trivial manner. One particular term of interest would be
the so called ``relaxation'' coupling to $(\dow_t + u^k\dow_k) "s^{ij}$ in the
stress tensor, analogue of which was required in relativistic fluids to salvage
causality. Muller, Israel and Stewart noted in~\cite{Israel:1979wp,
  Israel:1976tn, Muller:1967zza} that the causal structure of relativistic
hydrodynamics can be recovered by adding a second order term proportional to
$u^{\r}\N_{"r}"s^{"m"n}$ to the energy-momentum tensor. The associated transport
coefficient is known as the ``relaxation time''. For a non-relativistic fluid,
the corresponding term is related to $\bar\delta_{13}$ in \cref{tab:NC-t}. It
will be interesting to study the effect of this term on the Stokes' law. The
analysis however, is quite involved. We intend to come back to this study in a
future project.

Another prospective direction would be to compute the explicit form of these
second order transport coefficients using holography. The principles of
hydrodynamics and the second law of thermodynamics allow us to pen down the
constitutive relations of a fluid up to some unknown transport
coefficients. Details of these transport coefficients, depend on the particular
fluid in question and the details of the microscopic theory. Holography however,
allows us to compute these coefficients directly for a particular class of
fluids. For relativistic fluids, using the fluid/gravity
correspondence~\cite{Bhattacharyya:2008jc, Baier:2007ix}, transport coefficients
for a holographic plasma has been successfully computed
(see~\cite{Policastro:2001yc, Bhattacharyya:2008jc, Banerjee:2009ju,
  Astefanesei:2010dk, Baier:2007ix, Banerjee:2010zd, Erdmenger:2008rm,
  Banerjee:2008th, Dutta:2008gf} and references therein). For a first order
non-relativistic fluid, some progress in this direction has been made
in~\cite{Rangamani:2008gi}. We would like to set up a holographic model dual to
our null fluid construction and use it to compute the associated transport
coefficients. It would be interesting to see what this analysis has to tell us
about the second order transport coefficients talked about in this paper.

\acknowledgements

We would like to thank Suvankar Dutta for various useful discussions and
collaboration during the initial stages of this work. We would also like to
thank Jyotirmoy Bhattacharya and Ruth Gregory for helpful comments during the
course of this project. We are thankful to IISER Bhopal where this project was
initiated. AJ would like to thank the hospitality at IISER Pune and Perimeter
Institute where part of this work was done. Work of NB is partially supported by
a DST/SERB Ramanujan Fellowship. AJ is supported by the Durham Doctoral
Scholarship offered by Durham University. Finally, we thank the people of India
for their generous support for the basic sciences.

\appendix

\addbibresources{custom}
\makereferences 

\end{document}